\documentclass[12pt]{article}

\usepackage{amsmath, amsfonts, amsthm}
\usepackage{mathrsfs,amssymb}
\usepackage{bbm}
\usepackage{lipsum}
\usepackage{xcolor}
\usepackage[titletoc]{appendix}

\usepackage[authoryear,round]{natbib}
\usepackage[linkcolor=blue,citecolor=blue, colorlinks]{hyperref}

\date{}
\renewcommand{\ref}{\par\noindent\hangindent 12pt}
\newcommand{\noi}{\noindent}

\theoremstyle{definition}

\newtheorem{proposition}{Proposition}[section]

\numberwithin{equation}{section}

\begin{document}
\title{Endogenous Structural Transformation in Economic Development}

\author{Jistin Yifu Lin$^a$ and Haipeng Xing$^{b*}$}
\maketitle

\vspace{-0.5cm}
\noi\rule{6.5in}{0.4pt}\\
\noi {\bf Abstract}: 

This paper extends Xing's (2023abcd) optimal growth models of 
catching-up economies from the case of production function switching
to that of economic structure switching and argues how a country \
develops its economy by endogenous structural transformation 
and efficient resource allocation in a market mechanism.
To achieve this goal, the paper first summarizes three 
attributes of economic structures from the literature, namely, 
structurality, durationality, and transformality, and discuss 
their implications for methods of economic modeling. 
Then, with the common knowledge assumption, the paper 
extends Xing's (2023a) optimal growth model that is based on
production function switching and considers an extended
Ramsey model with endogenous structural transformation
in which the social planner chooses the optimal industrial 
structure, recource allocation with the chosen structure, 
and consumption to maximize the representative household's 
total utility subject to the resource constraint. 
The paper next establishes the mathematical underpinning of 
the static, dynamic, and switching equilibria. 
The Ramsey growth model and its equilibria are then 
extended to economies with complicated economic structures 
consisting of hierarchical production, 
technology adoption and innovation, infrastructure, and 
economic and political institutions. The paper concludes 
with a brief discussion of applications of the proposed 
methodology to economic development problems in other 
scenarios. 

\bigskip

\noi\rule{6.5in}{0.4pt}\\
\noi $^*$ Corresponding author

\noi $^a$ Institute of New Structural Economics, 
Peking University, Beijing, China.

\noi $^b$ Department of Applied Mathematics and Statistics, 
State University of New York, Stony Brook, NY 11794, USA.
Email: \texttt{haipeng.xing@stonybrook.edu}

We are grateful to Tony Addison, 
Jones Chad, Gene Grossman, 
Joseph Kaboski, Sejik Kim, Edmund Phelps, Richard Rogerson,
Thomas Sargent, Joel Sobel, Philip Ushchev, Shlomo Weber, 
Yi Wen, Adrian Wood, and Bo Zhang for their suggestion on 
an earlier version of the paper. We also thank participants
of webinars/seminars held by the CEPR-STEG programme, Moscow State
University, Seoul National University, University of Copenhagen, 
Peking University for their helpful comment.

\newpage
\tableofcontents
\newpage

\setcounter{section}{0}
\setcounter{page}{1}
\section{Introduction}

The goal of this paper is to develop a framework of modeling 
endogenous structural transformation and efficient resource 
allocation in a market economy during different development
stages. Following 
North (1981), we define ``structure'' as the characteristics 
of an economy, which are the basic determinants 
of economic activities with three attributes, namely, 
structurality, durationality, and transformality, on which 
we will elaborate in the next section.
\footnote{The literature has not yet provided a widely 
accepted definition of the economic structure of an economy. 
In this paper, economic structure refers to a collection 
of basic determinants of the composition and organization of 
production, consumption, distribution, exchange, and other 
economic activities in an economy. This mainly includes
production structures, technological structures, market 
structures, consumer preference or consumption structures, 
population structures, financial system structures, trade 
structures, and even institutional and cultural structures in
an economy.

In addition to a country's economic characteristics, 
economists consider noneconomic characteristics as 
determinants of economic activities. For instance, 
Kuznets (1966, p.437) assumes that ``the economic and many 
noneconomic characteristics of social structure are interrelated 
as both causes and effects," and illustrates ``some of the 
noneconomic characteristics associated with differences in 
economic development and structure between underdeveloped and
developed countries bear upon: (1) demographic patterns, 
(2) political structure, and (3) cultural aspects." 

In his book explaining the stability or change of 
economic structures in economic history, 
North (1981, p.1) explains ``structure" as follows. ``By 
`structure' I mean those characteristics of a society which
we believe to be the basic determinants of performance. Here 
I include the political and economic institutions, technology,
demography, and ideology of a society."}
``Structural transformation" refers to changes in the
basic determinants of the composition and organization of
economic activities during the process of economic development.
``Endogenous structural transformation" means that, for the 
purpose of economic development, the social planner in the 
economy makes use of knowledge on economic 
structures and chooses optimal economic structures to 
transform during economic development. 
Finally, ``from the early, catching-up stage to the 
advanced, sustained stage" means the process of how a 
less developed economy moves from inside the 
world production possibilities frontier toward its frontier.

To illustrate the idea and present our framework, 
we begin with the study of structural transformation and 
economic growth in the neoclassical sense. 
Structural transformation, or structural change, has been 
referred to as the process of a country's productive resources 
relocating from low-productivity to high-productivity 
economic activities during the past decades. 
Lewis (1954) argues  the importance of structural 
transformation from a dual economy to an industrialized 
market economy in the first stages of economic development, 
and Kuznets (1966, 1973) explains the economic growth as a sustained 
increase in per capita income accompanied by ``sweeping structural changes."
\footnote{Kuznets (1966, p.1) explains the economic growth 
as follows: ``We identify the economic growth of nations as 
a sustained increase in per capita or per worker product, 
most often accompanied by an increase in population and 
usually by sweeping 
structural changes. In modern times these were changes in the 
industrial structure within which product was turned out and 
resources employed–away from agriculture toward nonagricultural 
activities, the process of industrialization; in the distribution 
of population between the countryside and the cities, the process 
of urbanization; in the relative economic position of groups 
within the nation distinguished by employment status, attachment 
to various industries, level of per capita income, and the like; 
in the distribution of product by use–among household consumption, 
capital formation, and the government consumption, and within 
each of these major categories by further subdivisions; in the 
allocation of product by its origin within the nation's 
boundaries and elsewhere; and so on."}
Kuznets (1966, 1973) further documents the long-run 
transformation in several economies at the sector level, 
that is, the reallocation of economic resources from 
agriculture to the manufacturing and services sectors
\footnote{Kuznets (1973, p. 248) concludes that "major aspects 
of structural change include the shift away from agriculture 
to nonagricultural pursuits and, recently, away from industry 
to services."}
and lists it as one of the six major stylized factors
of long-term growth. 
\footnote{Stylized facts on sectoral structural transformation 
have been documented in the literature, which is too large to 
be entirely listed here. Sectoral structural transformation 
is usually demonstrated via sectoral shares of employment, 
value added, and final consumption expenditure; see a recent 
summary on this in Herrendorf, Rogerson, and Valentinyi 
(2014, section 2).}
To explain these sectoral reallocation processes, multisector 
models are developed to explore the linkage between sectoral 
structural transformation and balanced or non-balanced growth
and explain sectoral structural transformation as the result 
of changes in income and/or relative prices of goods.
\footnote{See Acemoglu and Guerrieri (2008), 
Herrendorf, Rogerson, and Valentinyi (2014), and the 
references therein.}
These multisector models generate results that are 
consistent with the ``stylized facts" of sectoral structural 
transformation and provide insights on several interesting 
economic issues, such as economic development, regional
income convergence, and aggregate productivity trends. 
However, they do not feature Kuznets' (1961) sweeping 
structural change, due to the following two reasons. 

First, Kuznets' sweeping structural change involves changes 
in basic determinants of the composition and organization 
of economic activities, such as agrarian production and 
other institutions related to modern industrialized production 
and related institutions. By contrast, sectoral structural
transformation in the multisector models deals only with 
variation in resource allocation among three given sectors 
under a specific economic structure.
\footnote{Acemoglu (2009, pp. 693-696) refers to ``structural 
change" as  changes in the composition of production and 
employment and uses the term ``structural transformation" 
to describe changes in the organization and efficiency of 
production accompanying the process of development. 
Acemoglu explains that, ``one might
expect Kuznets' structural change to be accompanied by a process 
that involves the organization of production becoming more 
efficient and the economy moving from the interior of the 
aggregate production possibilities set toward its frontier." 
Therefore, ``we would like to develop models that can account 
for both the structural changes and transformations at the
early stages of development and the behavior approximated by 
balanced growth at the later stages." }
Zooming in on structural transformation from the sector 
level to the industry level, changes in resource allocation
among the three given sectors---agriculture, industry, and
services---result from the transformation of industrial 
structures, and such transformation consists of the 
birth and decay of various industries during economic 
development, which the multisector growth models cannot 
describe.
\footnote{Ju, Lin and Wang (2015) provide an empirical study 
on this and further explain the phenomenon by an 
infinite-industry growth model that assumes geometrically 
distributed production functions and exogenous goods prices.} 
Second, Kuznets' sweeping structural change is referred to 
as structural transformation involving changes in the 
composition of industries; rural-urban population ratios; 
hard infrastructure; and social, economic, legal, and 
political institutions and so forth.
Multisector models only deal with sectoral structural
transformation in the process of economic growth
and hence fail to explain different kinds of structural 
transformation in the economic history of many countries 
in Europe, the Americas, Asia, and Africa. 
\footnote{Early efforts characterizing economic development 
in underdeveloped countries include Hirschman's (1958) emphasis 
on unbalanced growth, Nurkse's balanced growth, and
Rosenstein-Rodan's big push model. Dual economic models were 
developed to describe the process of structural changes 
that occurred in the early stages of economic development; 
see Lewis (1954), Ranis and Fei (1961), and others. }
Realizing the limits of the multisector growth models, 
some theoretical economists highlight the importance of 
characterizing ``sweeping structural changes" and advocate
for a unified theoretical framework to characterize the process
of structural transformation and resource allocation 
during economic development and growth.
\footnote{Acemoglu (2009, pp. 693-696) argues as follows:
``A useful theoretical perspective might therefore be to 
consider the early stages of economic development taking 
place in the midst of---or even via---structural changes 
and transformations. We may then expect these changes to 
ultimately bring the economy to the neighborhood of balanced 
growth, where our focus has so far been. If this perspective 
is indeed useful, then we would like to develop 
models that can account for both the structural changes and 
transformations at the early stages of development and the 
behavior approximated by balanced growth at the later stages."}
However, such a framework has never been developed. 

Assuming the agents in catching-up economies can decide
to choose production functions based on their backwardness
advantage, Xing (2023abcd) developed optimal growth 
models for catching-up economies that involve diminishing-
or constant-return-to-scale production functions and have
or don't have government spending. 
These models explain the distinction of the 
competitive equilibria between frontier and catching-up 
economies and characterize different scenarios of 
optimal growth paths of catching-up economies. 
Therefore, this paper extends the optimal growth 
models of catching-up economies in Xing (2023abcd) 
that are based on switching of production functions
to economies that involve switching of economic structures. 
we start with a summary of 
three attributes of economic structures and proceeds
step by step to develop a theoretical framework that
characterizes an economy's development and 
growth via {\it endogenous structural transformation} (EST) 
and effective resource allocation. Secton 2 argues that 
all structures possess three important attributes: 
structurality, durationality, and transformality. In addition
to illustrating and discussing these three attributes,
the section discusses the implications of the attributes
on methods of modeling structural transformation.

Section 3 summarizes and extends the Ramsey type model 
in catching-up economies studied in Xing (2023a), which assumes
that knowledge on the world's industrial structures can be 
freely obtained and a unique final good can be produced 
by different aggregate production functions. The economic 
structure of the economy is featured with a specific aggregate 
production function, and the social planner in the economy 
must optimally and dynamically choose an economic structure, 
resource allocation within the chosen structure,
and a consumption level over time to maximize the 
representative household's total utility subject to 
the endowment constraint. 

By using a decoupled representation of economic activities and 
their structures, the social planner's objective can be written 
as an infinite-horizon optimal control problems for hybrid 
dynamical systems. 
Xing (2023a) shows that the the value function in the 
social planner's maximization satisfies {\it Hamilton-Jacobi-Bellman 
equations and quasi-variational inequalities} (HJBQVI), and 
establishes the associated competitive equilibrium theory. 
We argue that the competitive equilibrium in Xing (2023a) 
actually consists of three types of equilibria, static, dynamic,
and {\it switching equilibrium}. While the former two equilibria
determine efficient resource allocation and imply optimal 
paths of economic activities in a given econoimc structure, the 
switching equilibrium characterizes optimal economic 
structures at each time period and hence their path.
In the special case that the economy has only a single 
economic structure to choose, the switching equlibrium 
degenerates so that the extended Ramsey model reduces to 
the neoclassical growth model. 
From this perspective, the extended Ramsey model extends 
the neoclassical economy from a single structure to 
multiple structures with possible transformation
among them. Section 4 further discusses 
the impacts of the factor endowments on the optimal structures 
and transformation regions of the factor endowments and
provides some examples and their implications for economic
development.

The optimal growth model in Xing (2023abcd) dealt with 
switching of production functions, but the idea of the 
model can be further extended and
applied to economies that have complex structures at 
different development stages. Section 5 illustrates this by 
describing EST in different types of structures and 
discussing their patterns of stagewise development and growth. 
The first type contains a class of hierarchical production 
structures with which intermediate and final goods are 
produced with exogenous technological progress. 
The second type deals with structures of endogenous 
technological progress achieved by adoption and/or 
{\it research and development} (R\&D) and their transformation. 
The third type deals with transformation of
economic institutions involving infrastructures and economic 
policies. The fourth type handles structural transformation 
of political institutions.

Section 6 provides a brief discussion and concluding remarks 
on insights from the EST and proposes areas for future research.
Given that the mechanism of EST in the current 
framework is based solely on the market, the section discusses the significance 
and necessity of a framework that integrates the market 
mechanism and government intervention. 

\section{Structurality, durationality, and transformality}

In the literature, an economy's structures usually consist 
of economic and noneconomic characteristics of activities. 
The former include a set of specifications on the composition 
and organization of production and other activities, and 
the latter consist of a collection of rules on economic 
and political institutions and even the norms, values, 
and ideologies of the society. Since these characteristics 
determine the ways in which economic activities are organized 
and managed, they are different from numerical economic  
variables that measure the levels of input and output of 
economic activities. In particular, three specific 
attributes are found in an economy's 
structures---structurality, durationality, and transformality.

\subsection{Illustration and discussion}

The first attribute is {\it structurality}, which refers to 
the organic relationships of all the economic and noneconomic 
structures, each with specific characteristics, in the overall 
structure of an economy or society. An economy's overall 
structure is a collection of all the structures, which are 
organized and interconnected according to certain rules. 
Different schools of thought have proposed different theories 
to explain the fundamental determinant of an economy's overall 
structure and the change, say, from an agrarian economy to a 
modern industrialized economy. For example, according to 
Weber (1904, 1905), it is value; according to Marx and 
Engels (1848), it is productive force; and according 
to North (1981), it is the land-to-labor ratio 
that determines the overall structure of a society.
\footnote{
In Marx's historical materialism, the superstructure, or 
institutions, is determined by and impacts the economic base, 
which consists of the productive forces and the production 
relations determined by the productive forces. The productive 
forces are determined by the prevailing technology and 
industries in an economy.}
In neoclassical economics, the competitive equilibrium 
suggests that, among all the endogenous variables, capital 
intensity (or generally, the factor endowments) determines 
efficient resource allocation, which includes production, 
consumption, and endogenous technological changes. 
In institutional economics, the role of institutions in 
shaping economic behavior is highlighted, and institutional 
structure seems to be more fundamental than the factor endowment. 
In new structural economics the endowment structure
determines the comparative advantages, that is, the industrial 
structure and appropriate hard infrastructure and soft 
institutions of an economy in the process of development. 
In the literature, the word ``structure" is sometimes used 
to refer to the overall structure or alternatively 
the totality of all (sub)structures in a society; at other 
times, it is used to refer to a specific (sub)structure 
in the overall structure. 
In the latter case, an adjective is often added, for example, 
the endowment structure, the industrial structure, the 
preference structure, the financial structure, the legal 
structure, the institutional structure, the political 
structure, and so forth.

The second attribute is {\it durationality}, which means 
that the overall structure and each of its substructures in 
an economy will not change instantaneously and will have 
different levels of stability. Compared with numerical 
economic variables that measure the input and output levels 
of economic activities, economic structures describe how economic 
activities are organized and managed and/or how economic 
variables are generated and, hence, do not change as fast as 
numerical economic variables. For instance, an economy's industrial 
structure does not change instantaneously with variation in
firms' output and technology levels. Similarly, a firm's 
structure, including its equipment and organization, does not 
change instantaneously with its output level.
Furthermore, different (sub)structures show different extents 
of durationality. One example of this is that quality 
improvement in Schumpeterian growth occurs during the 
lifetime of each variety of machine; hence, the characteristics 
(or structure) of improvement in the quality of each 
type of machine are less durable than the structure of 
production that determines whether a particular type of 
machine should be produced. Another example is that structures 
of political institutions in an economy are more durable 
than other kinds of economic (sub)structures, such as  
technological progress structures, industrial structures, 
and even structures of economic institution. In addition, the 
overall structure of an agrarian society and its related 
substructures, including economic, social, and political 
structures, may have existed for millennia in a society, 
and change of the overall structure to an industrialized 
society has taken place only in modern times.  

The third attribute is {\it transformality}, which means 
that a substructure or even the overall structure is not 
constant forever; it can transform into another structure 
under certain conditions
during the process of economic development and growth. 
Transformation of economic (sub)structures can be easily found 
in countries' development and growth processes and has been 
widely discussed by economists over the past centuries. Since the
literature documenting structural transformation is too 
large to be entirely reviewed here, we emphasize the
difference between the transformality of economic structures 
and the variability of numerical economic variables. In contrast
to numerical economic variables, which change almost instantaneously, 
economic structures change on a much longer time scales due to 
their durationality. Furthermore, economic (sub)structures 
have different degrees of transformality in economic development. 
For instance, it is typically much easier for a country to 
transform the structure of industry from labor intensive
to capital intensive and technological progress from 
technological borrowing to indigenous innovation
than to transform the institutional structures of the economy.

\subsection{Implication for the methodology of economic modeling}

The three attributes distinguish economic structures from 
economic variables of resource allocation and also 
from themselves, and such distinction indicates that 
economic structures should be characterized by a 
method different from that of modeling
economic variables. In the following, we briefly discuss 
several issues related to modeling economic structures.

\subsubsection*{Decoupling economic structures and 
economic variables}

Since a country's economic activities are measured by their input
and output levels and determined by their composition and
organization, their characterization can be decoupled into 
two types of variables. The first type is a set of functional 
variables that describe economic (sub)structures or determinants 
of economic activities, and the second type is a set of 
numerical variables that represent the inputs and outputs of 
economic activities. As an example of this, consider an 
economy that has a unique 
final good produced by aggregate production functions. Capital 
stock, labor, and output are the numerical variables that serve
as inputs and outputs of the production functions, and the 
aggregate production functions are functional variables that 
determine the production process and hence represent the 
production structure of the economy. If exogenous or 
endogenous technological progress is considered in the economy,
the structure of technological progress in production is 
represented by functionals that determine the technological 
progress. 

The decoupling idea separates economic structures from  
measurements in economic activities and allows us to deal with
functional and numerical variables differently in economic 
modeling. Specifically, given a functional variable (or an 
economic sub-structure) that determines specific economic 
activities, the dynamics of numerical economic variables can be
analyzed by existing economic methods. When an economic 
(sub)structure transforms from one to another, it may be 
considered as a change of the functional variables. Economic
(sub)structures with different extents of durationality
may be represented as functional variables on different
time scales. 

\subsubsection*{Structures of economic (sub)structures}

By representing each economic substructure as a functional
variable, a country's economic structure becomes a collection
of functional variables that determine various economic 
activities and are organized by certain economic rules. 
For instance, consider an economy that has intermediate 
and final goods. The production structure of the economy is 
represented as a functional that aggregates the production 
functions of the intermediate and final goods, and 
the structure of consumers' preference is described 
by a functional of intermediate (and/or final) goods. These
two substructures are aggregated as part of the 
economic structure of the economy under the resource constaint. 

However, the economic rule of organizing economic 
(sub)structures may not be unique, and sometimes the cause 
and effect of economic (sub)structures is controversial 
among economists. As exploration of such 
rules goes beyond the scope of this paper, we postulate
that the cause and effect of economic (sub)structures are
known by economists; hence, the structure of economic 
(sub)structures can be well specified by economic researchers. 
Once the cause and effect of economic (sub)structures 
are clear, the pivot or central structure of all the economic
(sub)structures may be identified.

\subsubsection*{knowledge on economic structures}

The transformation of economic structures indicates that 
more than one economic structure appears in a country's 
development and growth process. To describe the transformation 
process, we briefly discuss how knowledge on
economic structures is generated for a country. 
In principle, knowledge on economic structures is created or 
summarized by countries' political, economic, and intellectual
elites. Take the structure of technological progress as 
an example which refers to the organization of economic 
activities to improve an economy's technology level.
For countries inside the world production possibilities
frontier, their knowledge on structures of technological 
progress consists of not only ways to carry out R\&D by
themselves, but also the paths of other countries' 
technological progress for adoption. 
\footnote{A potentially confusing issue here is the difference 
between the technological structure and the structure of 
technological progress. As the former usually refers to the 
composition and level of technology in production, we refer to 
the latter as how the level of technology is determined 
in economic activities.}
However, this is not the case for countries on the world 
production possibilities frontier, since they may not need 
to know paths of technological progress in countries 
inside the world production possibilities frontier.
Thus, their knowledge on structures of technological 
progress only consists of R\&D and/or learning
by doing during their development process. 

Similar argument can be applied to knowledge on other 
kinds of structures. But for structures involving noneconomic 
characteristics, the knowledge on them may be obtained
in different ways.
Take the knowledge on structures of legal institutions 
as an example. Since each country's development, culture,
and social environment is unique, his/her knowledge on 
structures of legal institutions is also unique and may not
be useful for other countries.

\subsubsection*{Role of the social planner}

Once the knowledge on economic structures is obtained, the
next concern is about who will make use of the knowledge
and make decisions on the transformation of economic 
structures. It may be assumed that such decisions are made by interactions and joint efforts of the countries' political, 
economic, and intellectual elites at different levels. 
For example, when a new structure of technological progress
for a machine type needs to be adopted in countries inside the
global production possibilities frontier, an association of
entrepreneurs may make the decision; if such decision needs 
to be supported by workers with different skill sets trained 
in the schools or through subsidies provided by the government, 
the government will decide whether a new technology structure 
should be adopted and provide education and subsidies accordingly.
Decison makers for other types of economic substructures 
usually depend on countries' economic, political, and cultural
institutions and should be carefully studied. To model the idea, 
we do not discuss the details of this here and assume simply that 
the social planner can make decisions on the transformation of
economic structures.

\section{A Ramsey growth model with endogenous structural transformation}

This section extends Xing (2023a)'s optimal growth model 
of catching-up economies to the case in which the social 
planner makes decisions on resource allocation 
for production activities within an industrial structure 
to meet the households' consumption demands and 
transformation of industrial structure within the given 
overall structure of the economy. 

\subsection{Representation of economic activities and
their structures}

\subsubsection*{Households, firms, production, and technology}

Let $\mathcal{H}$ be the set of households in the economy
and suppose that the economy admits a representative 
household. That is, the demand side of the economy can 
be represented as if there were a single household making 
the aggregate consumption and saving decisions subject to 
an aggregate budget constraint. Denote this characteristic by 
$\mathscr{H}=\{ \mbox{households in } \mathcal{H} 
\mbox{ are representative} \}$.
The behavior of households in the economy is characterized
by the pair $(\mathcal{H}, \mathscr{H})$.

Similarly, let $\mathcal{F}$ be the set of firms in the 
economy and suppose that all firms are representative.
That is, all firms in the economy access the same aggregate
production function for the final good. Denote this
characteristic by $\mathscr{F} = \{ \mbox{firms in }
\mathcal{F} \mbox{ are representative}\}$. The behavior 
of firms in the economy is characterized by the pair 
$(\mathcal{F}, \mathscr{F})$. 

Suppose that firms use the aggregate production function
$Y(t)=F(K(t), L(t), A(t))$ to produce the final good, 
where $K(t)$, $L(t)$, and $A(t)$ are, respectively, the 
capital stock, employment, and technology used in production 
at time $t$. Let $\mathcal{Y}=\{ Y(t) \in \mathbbm{R}^+ 
\}$ represent the 
output of production and $\mathscr{Y}=\{ F\}$ represent 
the composition and organization of production. Then, 
$(\mathcal{Y}, \mathscr{Y})$ characterizes
production and its organizational structure.

The level of the technology $A$ is exogenously determined
by a function of technological progress, $A(t)$. 
$\mathcal{A}=\{ A(t) \in \mathbbm{R}^+ \}$ denotes the level of technology 
at time $t$ and $\mathscr{A}=\{ A: \mathbbm{R}^+
\rightarrow \mathbbm{R}^+$ be exogenously 
given.$\}$ represents how the technology is exogenously 
generated. Then, technology and its structure of production 
can be characterized by $(\mathcal{A}, \mathscr{A})$. 
$A(t)$ may depend on the production function.

\subsubsection*{Institution, labor, resource constraint,
and price}

To specify a way of allocating resources, we assume that 
all goods and factor markets are competitive and complete. 
Let $\mathcal{M}=\{$the demand and supply of labor, 
capital, and the goods$\}$ and $\mathscr{M}=\{$households
and firms are price-takers and pursue their own goals and
prices clear markets and all markets are complete.$\}$. 
Then the composition and institutional structure
of the market can be expressed as
$(\mathcal{M}, \mathscr{M})$.

Given $(\mathcal{M}, \mathscr{M})$, we explore the 
firms' demand for labor and capital in the aggregate 
production $(\mathcal{Y}, \mathscr{Y}_i)$ and their 
structures. Assume the population $\overline{L}$ grows 
exponentially at rate $\pi$ $(\pi>0)$, that is,
\begin{equation}\label{sec3.popu1}
\overline{L}(t): = \overline{L}_{\pi} (t) = 
\overline{L}(0) \exp (\pi t), \qquad \pi > 0.
\end{equation}

\noi Hence, the population structure is expressed as
$(\overline{\mathcal{L}}, \overline{\mathscr{L}})
=( \{ \overline{L}(t) \}, \{ \overline{L}_\pi \})$,
where $\overline{L}_\pi$ represents the functional 
\eqref{sec3.popu1}. To describe the labor market, 
note that $L(t) \in \mathbbm{R}^+ $ is the amount of demand for labor and 
$L(\cdot)$ is a functional characterizing such demand;
hence, the labor market and its structure are 
represented by $(\mathcal{L}, \mathscr{L})
=( \{ L(t) \}, \{ L(\cdot) \})$. When the market 
is competitive, the labor market-clearing condition 
is $L(t)=\overline{L}(t)$, suggesting that 
the labor market structure coincides with the 
population structure, that is, $(\mathcal{L}, \mathscr{L}) 
\cong (\overline{\mathcal{L}}, \overline{\mathscr{L}})$.
\footnote{This further implies that, if there
exist two growth rates $\pi_1>0, \pi_2>0$, 
$\pi_1\neq \pi_2$, then $( \{ L(t) \}, \{ L_{\pi_1} \})$ 
and $(\{ L(t) \}, \{ L_{\pi_2} \})$ are two different 
representation of the labor market in the economy.}

The households own the capital stock of the economy and 
rent it to firms. Under the capital market-clearing 
condition, the demand for capital by firms equals 
the supply of capital by households, which is denoted
$K(t)$. The aggregate resource constraint, 
which is equivalent to the budget constraint of the 
representative household, requires that
\begin{equation}\label{sec3.aggre.capital.equ1a}
\dot{K}(t) = Y(t) - \delta K(t) - C(t),
\end{equation}

\noi where $C(t):= c(t) L(t)$ is the total consumption, 
and investment consists of new capital $\dot{K}(t)$, and 
replenishment of depreciated capital $\delta K(t)$.
The constraint \eqref{sec3.aggre.capital.equ1a}
suggests that the capital market and its
structure can be characterized by $(\mathcal{K}, 
\mathscr{K})$, in which $\mathcal{K}=\{ K(t) \in 
\mathbbm{R}^+ \}$, 
and $\mathscr{K}=\{ K(\cdot) | K(\cdot) \mbox{ satisfies 
\eqref{sec3.aggre.capital.equ1a}} \}$.

Under appropriate assumptions (or specifically, 
(A1) and (A2) in section 3.2), the factor prices can
be obtained by solving the profit maximization problem 
of the representative firm. Denote by $R(t)$ and
$w(t)$ the rental price of capital and wage at
time $t$, respectively. Let $P(t):=1$ be the normalized
price of the final good. Factor and goods prices and 
their structures can be expressed as $(\mathcal{P}, 
\mathscr{P})$, in which $\mathcal{P}=\{ R(t), w(t), 
P(t) \}$ and $\mathscr{P}=\{$factor prices are 
determined by firms to maximize their profit, and 
$P(t)=1 \}$. Furthermore, the rental price and wage
under $(\mathcal{P}, \mathscr{P})$ can be expressed as
$R(t)=\frac{\partial F}{\partial K}(K, L, A)$
and $w(t)=\frac{\partial F}{\partial L}(K, L, A)$, respectively.

\subsubsection*{Consumption and utility function}

We now consider consumption in the economy and its
characteristic. Suppose that households only consume
the unique final good. Then total consumption is given 
by $C(t)=c(t) L(t)$ and the characteristic of the 
consumption in this economy is $\mathscr{C}=\{$households 
consume the final good$\}$.
Therefore, consumption and its characteristic are
expressed as $(\mathcal{C}, \mathscr{C})$. Given the 
consumption per capita $c(t)$, we assume that the 
representative household has an instantaneous utility 
function $u(c)$, which represents the preference of the 
household's consumption. Then, the preference of
consumption and its structure can be expressed as 
$(\mathcal{U}, \mathscr{U}):=(\{ u(c) \}, \{ u(\cdot) \})$.

\subsubsection*{Economic structure}

The above argument describes the following agents' behavior 
and economic activity components in the economy and 
their characteristics, 
$(\mathcal{H}, \mathscr{H})$, 
$(\mathcal{F}, \mathscr{F})$,
$(\mathcal{Y}, \mathscr{Y})$,
$(\mathcal{M}, \mathscr{M})$,
$(\mathcal{L}, \mathscr{L})$,
$(\mathcal{K}, \mathscr{K})$,
$(\mathcal{A}, \mathscr{A})$,
$(\mathcal{P}, \mathscr{P})$,
$(\mathcal{C}, \mathscr{C})$, and 
$(\mathcal{U}, \mathscr{U})$. 
To represent these in a more concise way, we let 
$\mathcal{E}:=( \mathcal{H}, \mathcal{F}, \mathcal{M}, 
\mathcal{Y}, \mathcal{L}, \mathcal{K}, \mathcal{A},
\mathcal{P}, \mathcal{C}, \mathcal{U})$ represent
agents' behavior and economic activities in the economy, 
and $\mathscr{E}:=( \mathscr{H}, \mathscr{F}, \mathscr{M}, 
\mathscr{Y}, \mathscr{L}, \mathscr{K}, \mathscr{A},
\mathscr{P}, \mathscr{C}, \mathscr{U})$ represent
the economic and noneconomic characteristics of
$\mathcal{E}$. Then the overall structure and activities 
of the economy can be characterized 
by $(\mathcal{E}, \mathscr{E})$. 

\subsection{Knowledge on economic structures}

Suppose that the final good in the economy can be produced
by $I$ aggregate production functions or industrial 
structures in the world:
\begin{equation}\label{sec3.prod1}
Y(t) = F_i[K(t), L(t), A(t)], 
\qquad i \in \mathbbm{I}:=\{1, \dots, I \}, 
\end{equation}

\noi and knowledge on these industrial structures 
can be freely obtained by economies in the world. 
The knowledge on the world's industrial 
structures for the social planner is then given by 
$\mathcal{I}_{\mathscr{Y}}=\{ \mathscr{Y}_i |
\mathscr{Y}_i=\{ F_i \}, i\in \mathbbm{I} \}$. 
When the social planner of the economy chooses 
$F_i$ to produce the final good, the industrial 
structure of the economy is described by the pair 
$(\mathcal{Y}, \mathscr{Y}_i) =(\{ Y(t) \}, \{ F_i \})$, 
and correspondingly, the economy with industrial 
structure $\mathscr{Y}_i$ in its overall structure is
characterized by $(\mathcal{E}, \mathscr{E}_i)$.
The knowledge on the industrial structures
for the social planner can now be expressed as 
$\mathcal{I}_{\mathscr{E}} :=\{ \mathscr{E}_i | 
i \in \mathbbm{I} \}$.

Provided the economy and its economic structure $(\mathcal{E}, 
\mathscr{E}_i)$, equations \eqref{sec3.prod1} and 
\eqref{sec3.aggre.capital.equ1a} imply the following
aggregate resource constraint in the economy
\begin{equation}\label{sec3.aggre.capital.equ1}
\dot{K}(t) = F_i[K(t), L(t), A(t)] - \delta K(t) - C(t).
\end{equation}

\noi Suppose that the production function $F_i[K, L, A]$
exhibits constant returns to scale in $K$ and $L$. The
output per capita is given by 
\begin{equation}\label{sec3.per.prod1}
y(t) \equiv Y(t)/L(t)  \equiv f_i(t, k(t)), \qquad
f_i(t, k(t)) = F_i [k(t), 1, A(t)],
\end{equation}

\noi where $k(t) \equiv K(t)/L(t)$. Then, the accumulated
capital per capita is given by the equation
\begin{equation}\label{sec3.per.capital.equ1}
\dot{k}(t) = f_i(t, k(t)) - (\delta+ \pi) k(t) - c(t). 
\end{equation}

\noi We assume that $f_i$ $(i \in \mathbbm{I})$ satisfy
the following conditions for later discussion:

\medskip
(A1) {\it For each $i \in I$, the production function 
$f_i(t,k)$ is twice differentiable, strictly 
increasing, and concave in $k$.}

(A2) {\it For each $i$, $f_i$ satisfies the Inada 
conditions: $\lim_{k \rightarrow 0} \partial f_i(t, k) 
/\partial k = +\infty$ and $\lim_{k \rightarrow +\infty} 
\partial f_i(t, k) /\partial k = 0$.
Moreover, $f_i(t, 0)=0$ for all $t$.}
\medskip

\noi Furthermore, we assume that the utility function 
$u(c)$ satisfies the following condition:

\medskip
(A3) {\it The utility function $u: \mathbbm{R}^+ \rightarrow
\mathbbm{R}$ is strictly increasing, concave,
and twice differentiable, with derivatives $u'(c)>0$
and $u''(c)<0$ for all $c$ in the interior of its 
domain.}

\medskip

\subsection{Decisions on structural transformation
and resource allocation}

At each time $t$, the social planner determines an
economic structure (or more precisely, an industrial
structure) for the economy and allocate resources under 
the given overall structure. Let $\theta(t)$ be the 
economic structure (or with a slight abuse of 
notation, industrial structure) chosen 
at time $t$ by the social planner from the information 
set of economic structures $\mathcal{I}_{\mathscr{E}}=\{ 
\mathscr{E}_i | i \in \mathbbm{I} \}$, and denote by 
$e(t) \in \mathcal{E}$ households' and firms' behaviors 
and economic activities at time $t$. Then, given 
the industrial structure prior to time $t$, $\theta(t-)$,
the social planner decides whether the industrial structure
$\theta(t)$ should be the same as the previous one. If yes,
then $\theta(t)=\theta(t-)$; otherwise, the social 
planner chooses an industrial structure $\theta(t)$ $(\neq
\theta(t-))$ from $\mathcal{I}_{\mathscr{E}}$ 
and transforms the economic structure from $\theta(t-)$
to $\theta(t)$. After choosing the industrial structure, 
the social planner will allocate resouces and decide
the level of $e(t)$. Since all the economic structures in 
$\mathcal{I}_{\mathscr{E}}$ are different only in industrial
structures $\mathcal{I}_{\mathscr{Y}}$, $e(t)$ can be 
expressed as a vector of numerical economic variables 
$(k(t), R(t), A(t), w(t), c(t))$.

The above argument indicates that the path of
industrial structures $\{ \theta(t) \}$ is piecewise
constant and satisfies the durationality and 
transformality of structures discussed in 
section 2. Then, to present the idea, we represent 
$\{\theta(t) \}$ in another way. Let $\tau_n$ denote the 
time of the $n$th structural transformation and 
$\kappa_n \in \mathcal{I}_{\mathscr{E}}$ denote
the transformed structure at time $\tau_n$, for 
$n\in \{1, 2, \dots, \}$. Assume that this decision 
problem starts at time $t_0$, and let $\tau_0 = t_0$ 
and $\kappa_0=\mathscr{E}_i \in \mathcal{I}_{\mathscr{E}}$ 
(or $i \in \mathbbm{I}$). The series of $\tau_n$
and $\kappa_n$ satisfy

\medskip
(A4) $t_0=\tau_0 \le \tau_1 < \tau_2 < \cdots < 
\tau_n < \cdots$, and $\lim_{n\rightarrow +\infty} 
\tau_n = +\infty$. 
$\kappa_n \in \mathcal{I}_{\mathscr{E}}$ for all $n\ge 0$. 

\medskip
\noi In (A4), $\tau_1 \ge \tau_0$ means
that the social planner may decide to transform the 
industrial structure at the starting time $t_0$. 
Let $\xi = \{ (\tau_n, \kappa_n)_{n\ge 0} \}$ be 
the double series of transformation time and transformed
structures, and denote by $\mathfrak{A}$ the set of 
all series $\xi$ satisfying (A4). 
Then given an $\xi\in \mathfrak{A}$, the industrial 
structure  of the economy at time $t$ is expressed as
$$
\theta(t) = \sum_{n\ge 0} \kappa_n 1_{ [\tau_n, 
\tau_{n+1}) } (t), \quad \mbox{for } t \ge \tau_0=t_0,
$$

\noi where $1_{ [\tau_n, \tau_{n+1}) }(t)$ is an indicator 
function of $t$, taking value 1 if $t \in [\tau_n, 
\tau_{n+1})$ and 0 otherwise. By definition, $\theta(t)$ 
are right continous and have left limits at each $\tau_n$, 
and are piecewise constant over time $t$.

\subsection{Social planner's objective}

Suppose there is no market failure during the structural 
transformation, so that the social planner does not need to 
intervene the transformation. Then, given an $\xi = \{ 
(\tau_n, \kappa_n)_{n\ge 0} \} \in \mathfrak{A}$, the 
capital stock $K(\cdot)$ is continuous at transformation 
times $\tau_n$, that is, $K(\tau_n-)=K(\tau_n)$. 
Combining this with \eqref{sec3.aggre.capital.equ1} yields 
the capital accumulation process 
\begin{equation}\label{sec3.aggre.capital.equ2}
\left\{ \begin{array}{ll}
\dot{K}(t) =  F_{\kappa_n}[K(t), L(t), A(t)]- \delta K(t) 
- C(t), & \tau_n \le t < \tau_{n+1}, \\
K(\tau_n-) = K(\tau_n), &
t=\tau_n,  n=1, 2, \dots,
\end{array} \right.
\end{equation}

\noi Accordingly, the capital accumulation process per 
capita is expressed as
\begin{equation}\label{sec3.per.capital.equ3}
\left\{ \begin{array}{ll}
\dot{k}(t) = f_{\kappa_n}(t, k(t)) - (\delta+\pi) k(t) - c(t), & 
\tau_n \le t < \tau_{n+1}, \\
k(\tau_n-) = k(\tau_n), &
t=\tau_{n+1}, n=1, 2, \dots
\end{array} \right.
\end{equation}

\noi Thus, provided the initial overall 
structure $\theta(t_0)=\mathscr{E}_i$, the initial level 
of capital intensity $k(t_0)=k$, a path of structural 
transformation $\xi=\{ (\tau_n, \kappa_n)_{n \ge 0}\}$, 
and a path of consumption $\{ c(t) \}_{t\ge 0}$, the 
total utility for the representative household
starting at time $t_0$ is expressed as
\begin{equation}\label{sec3.totalutil.equ1}
J_i(t_0, k; \{c(t), \xi \} )=\int_{t_0}^\infty 
e^{-(\rho-\pi) t} u(c(t)) dt. 
\end{equation}

\noi The social planner's objective is to solve the 
maximization problem 
\begin{equation}\label{sec3.totalutil.equ2}
\begin{aligned}
& V_i(t_0, k) = \max_{ \{ c(t), \xi\} }  
J_i(t_0, k; \{c(t), \xi \} ) \\
& \qquad
\qquad \mbox{subject to \eqref{sec3.per.capital.equ3}} 
\mbox{  and } k(t_0) = k\in \mathbbm{R}^+, 
\theta(t_0)=\mathscr{E}_i \in \mathcal{I}_{\mathscr{E}}.
\end{aligned}
\end{equation}

\noi Since markets are complete and competitive, 
given an initial industrial structure $\theta(t_0)=i$
and an initial capital intensity $k(t_0)=k$, 
the competitive equilibrium is defined as the paths of
structures and the amount of consumption and savings 
$\{ \theta(t), c(t), \dot{k}(t) \}_{t\ge t_0}$
that maximize the household's total utility 
\eqref{sec3.totalutil.equ1} subject to the constraint 
\eqref{sec3.per.capital.equ3}.

\section{Competitive equilibrium}

We analyze the competitive equilibrium of the proposed 
model. Since Xing (2023ab) has
characterized all scenarios of optimal growth and the
competitive equilibrium qualitatively and numerically
for production function switching, we focus theoretical
characterization of the competitive equilibrium.

\subsection{Solution of the extended Ramsey model}

Note that the competitive equilibrium in the extended Ramsey model
consists of paths of consumption, capital stock,
wage rates, rental rates of capital, and industrial
structures, such that (i) the social planner maximizes the 
representative household's utility given the initial economic 
structure and initial capital intensity, wage rates and 
rental rates, and (ii) the paths of wage rates and rental rates
are such that given the paths of capital stock and labor, 
all markets clear. Using the method in Xing (2021) and
Xing (2023a), we can show the following. 

\begin{proposition}\label{sec5.nonstat.prop1}
{\it For each $i\in \mathbbm{I}$, $V_i(t,k)$ defined by 
\eqref{sec3.totalutil.equ2} is a viscosity solution to
\begin{equation}\label{sec5.hjbqvi.def1}
\begin{aligned}
& \max\Big\{ \sup_{c\in \mathfrak{U}} \Big[ 
\frac{\partial V_i}{\partial t}(t,k) + \big[ f_i(t,k)-
(\delta+\pi)k -c \big] \frac{\partial V_i}{\partial k}(t,k) 
+ e^{-(\rho-\pi)t} u(c) \Big], \\
& \hspace{100pt} \max_{j\neq i} V_j(t,k) -V_i(t,k) 
\Big\} = 0,
\end{aligned}
\end{equation}

\noi and such solutions are unique on $[0, \infty)\times 
\mathbbm{R}^+$.}
\end{proposition}

The above HJB-QVI system contains two components and their 
economic interpretation is clear. The first component 
determines optimal consumption and characterizes the 
transitional dynamics of the capital intensity and optimal 
consumption, given that the 
current economic structure $\mathscr{E}_i$ (or the 
current production structure $\mathscr{Y}_i$) is optimal. 
The second component compares the value functions associated 
with each of the economic structures and chooses the optimal 
one. In particular, given economic structure $\mathscr{E}_i$ 
prior to time $t$ and another 
economic structure $\mathscr{E}_j$ $(j\neq i)$, 
the social planner compares the total utility associated 
with $i$ and the total utility after transforming
instantaneously  from $\mathscr{E}_i$  to $\mathscr{E}_j$.
If the former is larger for any $j \in \mathbbm{I}$ and 
$j \neq i$, the economy should stay with economic 
structure $\mathscr{E}_i$ at time $t$ and then use the 
HJB component to determine the resource allocation within 
the given industrial structure and optimal consumption. 
Otherwise, the social planner should transform the
industrial structure and overall economic structure 
from $\mathscr{E}_i$  to $\mathscr{E}_j$ at time $t$. 

Proposition \ref{sec5.nonstat.prop1} has also the following
economic implication for some special cases of 
$\mathcal{I}_{\mathscr{E}} =\{ \mathscr{E}_1, \dots, 
\mathscr{E}_I \}$. When $I=1$ or 
$\mathcal{I}_{\mathscr{E}}=\{ \mathscr{E}_1 \}$, that is, 
there is only one economic structure available to the 
social planner, hence $i \equiv 1$ and the second 
component of HJB-QVI equation \eqref{sec5.hjbqvi.def1} 
is gone. Then \eqref{sec5.hjbqvi.def1} degenerates to 
the following HJB equation:
\begin{equation}\label{sec5.hjbqvi.I1}
\sup_{c\in \mathfrak{U}} \Big[ 
\frac{\partial V_1}{\partial t}(t,k) + \big[ f_1(t,k)-
(\delta+\pi)k -c \big] \frac{\partial V_1}{\partial k}(t,k) 
+ e^{-(\rho-\pi)t} u(c) \Big] = 0.
\end{equation}

\noi Equation \eqref{sec5.hjbqvi.I1} is the same as the 
standard HJB equation in the neoclassical growth model. Hence,
the competitive equilibrium associated with HJB-QVI system 
\eqref{sec5.hjbqvi.def1} with $I=1$ is the same as the one
associated with the neoclassical growth model.
When $\mathbbm{I}=\{ 1, 2\}$ or $\mathcal{I}_{\mathscr{E}}
=\{ \mathscr{E}_1, \mathscr{E}_2 \}$, HJB-QVI equations 
\eqref{sec5.hjbqvi.def1} hold for $i, j=1, 2, i \neq j$, and
the social planner of the economy
only needs to decide when to transform from the current
industrial structure and, thus, the economic structure, to 
the other one. When $\mathbbm{I}=\{ 1, \dots, I\}$ and 
$I\ge 3$, the social planner needs to choose
not only when but also which industrial structure 
to transform. The latter choice involves comparing 
two industrial structures other than the current one.

\subsection{Static, dynamic, and switching equilibria}

Proposition \ref{sec5.nonstat.prop1} implies that there
exist three types of equilibria in the EST model---static,
dynamic, and switching. We now
discuss these equilibria and their implications.

\subsubsection*{Static and dynamic equilibria and the Euler 
equation of consumption}

The first two types of equilibria are the static and 
dynamic equilibria when the economic structure 
$\mathscr{E}_i$ prior to time $t$ is still optimal at
time $t$. The static and dynamic equilibria
are characterized by 
\begin{equation}\label{sec5.statdyn.equil.equ1}
\left\{ \begin{array}{l} 
\sup_{c\in \mathfrak{U}} \Big[ \displaystyle
\frac{\partial V_i}{\partial t}(t,k) + \big[ f_i(t,k)-
(\delta+\pi)k -c \big] \frac{\partial V_i}{\partial k}(t,k) 
+ e^{-(\rho-\pi)t} u(c) \Big] = 0, \\
\max_{j\neq i} V_j(t,k) -V_i(t,k) < 0.
\end{array} \right.
\end{equation}

\noi The second component, $\sup_{j\neq i} V_j(t,k) -
V_i(t,k) < 0$, in \eqref{sec5.statdyn.equil.equ1} indicates 
that there does not exist an economic structure that is better 
than the current economic structure $\mathscr{E}_i$. 
Consequently, economic structure $\mathscr{E}_i$
(or production structure $\mathscr{Y}_i$) prior to 
time $t$ is still optimal at time $t$. Then,
$\theta(t-)=\theta(t) = \mathscr{E}_i$ and the equation 
\begin{equation}\label{sec5.stat.equil.equ1}
\sup_{c\in \mathfrak{U}} \Big[ 
\frac{\partial V_i}{\partial t}(t,k) + \big[ f_i(t,k)-
(\delta+\pi)k -c \big] \frac{\partial V_i}{\partial k}(t,k) 
+ e^{-(\rho-\pi)t} u(c) \Big] = 0
\end{equation}

\noi characterizes the static and dynamic equilibria 
that are associated with the current economic structure 
$\mathscr{E}_i$ and, hence, optimal consumption can
be determined. The first-order condition of 
\eqref{sec5.stat.equil.equ1} implies that optimal
consumption $c_i^*(t,k)$ solves the equation
\begin{equation}\label{sec5.stat.equil.equ2} 
e^{-(\rho-\pi)t} u(c_i^*(t,k))=\frac{\partial V_i}{
\partial k}(t,k).
\end{equation}

\noi Since $c_i^*(t,k)$ is the optimal consumption
associated with economic structure $\mathscr{E}_i$, 
optimal consumption $c_i^*(t,k)$ and $c_j^*(t,k)$ 
under two different economic structures $\mathscr{E}_i$
and $\mathscr{E}_j$ $(i\neq j)$ are usually different. 

To study the dynamics of optimal consumption, we first 
take the derivative of equation \eqref{sec5.stat.equil.equ2} 
with respect to $k$ and compare the result with 
\eqref{sec5.stat.equil.equ2}. Then we have
\begin{equation}\label{sec5.stat.equil.equ2a}
\dot{k} \cdot \frac{\partial^2 V_i}{\partial k^2}(t,k) =
- \epsilon_u(c_i^*) \cdot 
\frac{1}{c_i^*} \frac{\partial c_i^*}{\partial t}(t,k) \cdot
\frac{\partial V}{\partial k}(t,k).
\end{equation}

\noi where $\epsilon_u(c)= - u''(c)\cdot c/ u'(c)$ 
is the elasticity of the marginal utility $u'(c)$. 
Then taking the derivative of equation 
\begin{equation}\label{sec5.stat.equil.equ1a}
\frac{\partial V_i}{\partial t}(t,k) + \big[ f_i(t,k)-
(\delta+\pi)k -c_i^*(t,k) \big] \frac{\partial V_i
}{\partial k}(t,k) + e^{-(\rho-\pi)t} u(c_i^*(t,k)) = 0
\end{equation}

\noi with respect to $k$ and simplifying the result by
\eqref{sec5.stat.equil.equ2a}, we obtain the {\it 
Euler equation of consumption} associated with
economic structure $\mathscr{E}_i$:
\begin{equation}\label{sec5.stat.equil.equ6}
\frac{1}{c_i^*} \frac{\partial c_i^*}{\partial t}(t,k) 
= \frac{1}{\epsilon_u(c_i^*)}
\Big[ \frac{\partial f_i}{\partial k}(t,k) - \delta - \pi 
- \frac{\partial^2 V_i}{\partial t \partial k}
\Big/ \frac{\partial V_i}{ \partial k}\Big].
\end{equation}

\noi This can be viewed as a nonstationary version of
the Euler equation in the neoclassical growth models;
see Acemoglu (2009, section 8.2.2). Since assumption
(A3) suggests that the utility function is monotonically
increasing with $c$, the procedure of maximizing the total
utility by choosing economic structures can be considered
a procedure of comparing \eqref{sec5.stat.equil.equ6}
for different economic structures. We will elaborate on
this in section 4.4.

\medskip
Competitive factor markets imply that, when the economic 
structure at time $t$ is $\mathscr{E}_i$, the rental rate 
of capital $R_i(t)$ and the wage rate $w_i(t)$ are given by
\begin{equation}\label{sec5.stat.equil.equ3} 
R_i(t,k) = \frac{\partial F_i}{\partial K}[ K(t), L(t), A(t)] 
= \frac{\partial F_i}{\partial k}[ k(t), 1, A(t)]
= \frac{\partial f_i}{\partial k} (t,k),
\end{equation}

\noi and 
\begin{equation}\label{sec5.stat.equil.equ4} 
w_i(t,k) = \frac{\partial F_i}{\partial L}[ K(t), L(t), A(t)]
= f_i(t, k(t)) - k(t)\frac{\partial f_i}{\partial k} (t,k).
\end{equation}

\noi Given economic structure $\mathscr{E}_i$ and 
optimal consumption \eqref{sec5.stat.equil.equ2}, 
the dynamic equilibrium of the capital-labor ratio and
optimal consumption is characterized by the equation
\begin{equation}\label{sec5.stat.equil.equ5} 
\dot{k}(t) = f_i(t, k(t)) - (\delta+\pi) k(t) - c_i^*(t, k(t)).
\end{equation}

The static and dynamic equilibria in the EST model are 
not exactly the same as those in the neoclassical growth 
models. In the neoclassical growth models, the economic 
structure is fixed with $\mathscr{E}_i$, and hence, static 
and dynamic equilibria are defined whether the current 
economic structure $\mathscr{E}_i$ is optimal or not. 
By contrast, in the EST model, the static and dynamic
equilibria at time $t$ depend on the optimal 
economic structure $\mathscr{E}_i$ at that time. That is,
if an economic structure $\mathscr{E}_i$ is not optimal
at time $t$ and hence is not chosen by the social planner
of the economy, static and dynamic equilibria 
associated with economic structure $\mathscr{E}_i$ 
at time $t$ do not exist.

\subsubsection*{Switching equilibria and optimal 
industrial structures}

HJB-QVI system \eqref{sec5.hjbqvi.def1} also characterizes
a third type of equilibrium, which we refer to as 
{\it the switching equilibrium}. Under such an 
equilibrium,
\begin{equation}\label{sec5.struct.equil.equ1}
\left\{ \begin{array}{l} 
\sup_{c\in \mathfrak{U}} \Big[ \displaystyle
\frac{\partial V_i}{\partial t}(t,k) + \big[ f_i(t,k)-
(\delta+\pi)k -c \big] \frac{\partial V_i}{\partial k}(t,k) 
+ e^{-(\rho-\pi)t} u(c) \Big]<0, \\
\max_{j\neq i} V_j(t,k) -V_i(t,k) = 0.
\end{array} \right.
\end{equation}

\noi The HJB inequality in \eqref{sec5.struct.equil.equ1}
implies that, if the current economic structure were 
$\mathscr{E}_i$, the associated economy would not 
attain the static and dynamic equilibria no matter 
how consumption was chosen. Mathematically, the optimality 
principle fails when the economy is associated with 
the industrial structure embodied in economic structure 
$\mathscr{E}_i$. In contrast to this, the equality in 
\eqref{sec5.struct.equil.equ1} shows that there exists
another industrial structure $\mathscr{Y}_j$ (or equivalently,
an economic structure $\mathscr{E}_j \neq 
\mathscr{E}_i$) such that its associated value function 
(i.e., the maximized total utility) is greater than 
that associated with economic structure $\mathscr{E}_i$. 
Denoting the new optimal economic structure by 
$\mathscr{E}_{j^*(t,k)}$, $j^*$ satisfies the 
following condition:
\begin{equation}\label{sec5.struct.equil.equ2}
j^*(t,k)= \arg\max_{j \in \mathbbm{I}, j\neq i} V_j(t,k), 
\qquad V_{j^*}(t,k) = V_i(t,k)
\end{equation}

\noi This suggests that the optimal 
economic structure at time $t$ is expressed as
\begin{equation}\label{sec5.opt.struct.sol}
\theta(t) = \left\{ \begin{array}{ll}
\mathscr{E}_i & \mbox{if equation} \eqref{sec5.statdyn.equil.equ1} \mbox{ holds}, \\
\mathscr{E}_{j^*(t,k)} & \mbox{if equation} \eqref{sec5.struct.equil.equ1} \mbox{ holds},
\end{array} \right.
\end{equation}

\noi in which $j^*(t,k)$ is given by condition \eqref{sec5.struct.equil.equ2}. 

The discussion above assumes positive switching costs
that satisfy assumption (B3). In the degenerate case
of vanishing switching costs $\eta_{ij}(t) \equiv 0$, 
the discussion in Section 4.5 implies that the value 
function $V_1(t,k)=\dots=V_I(t,k)=V(t,k)$
is a solution of 
\begin{equation}\label{sec5.struct.equil.equ4}
\frac{\partial V}{\partial t}(t,k) + 
\max_{i \in \mathbbm{I}} \Big\{ \sup_{c\in \mathfrak{U}} 
\Big[ \displaystyle \big( f_i(t,k)-
(\delta+\pi)k -c \big) \frac{\partial V}{\partial k}(t,k) 
+ e^{-(\rho-\pi)t} u(c) \Big]\Big\}=0.
\end{equation}

\noi This suggests that the optimal economic structure at 
time $t$ is given by $\theta(t)= \mathscr{E}_{j^*(t,k)}$,
in which $j^*(t,k)$ satisfies the following two
conditions:
\begin{equation}\label{sec5.struct.equil.equ5}
j^*(t,k) = \arg\max_{i \in \mathbbm{I}} \Big\{ 
\frac{\partial V}{\partial t}(t,k) + \sup_{c\in 
\mathfrak{U}} \Big[ \displaystyle \big( f_i(t,k)-
(\delta+\pi)k -c \big) \frac{\partial V}{\partial k}(t,k) 
+ e^{-(\rho-\pi)t} u(c) \Big] \Big\},
\end{equation}

\noi and 
\begin{equation}\label{sec5.struct.equil.equ6}
\frac{\partial V}{\partial t}(t,k) + \sup_{c\in 
\mathfrak{U}} \Big[ \displaystyle \big( f_{j^*}(t,k)-
(\delta+\pi)k -c \big) \frac{\partial V}{
\partial k}(t,k) + e^{-(\rho-\pi)t} u(c) \Big] \Big\}=0.
\end{equation}

\medskip
The discussion above implies the following property:

\begin{proposition}\label{sec5.nonstat.prop2}
{\it The optimal industrial structure in an overall structure 
at any given time $t$ and with any given capital intensity
$k$ is determined by switching equilibrium
\eqref{sec5.hjbqvi.def1} at $(t,k)$. Moreover, 
the optimal industrial structure is a function of $t$ 
and $k(t)$, and hence endogenous to the capital 
intensity (or more generally, the factor
endowments of the economy).}
\end{proposition}

\noi Furthermore, since $\mathscr{E}_{j^*(t,k)}$ is the optimal 
industrial structure for the given $(t,k)$, the social 
planner should transform the economic structure of the 
economy from $\mathscr{E}_i$ to $\mathscr{E}_{j^*(t,k)}$
at $t$. Consequently, the optimal consumption,
rental rate of capital, and wage rate will 
shift from $c_i^*(t,k)$, $R_i(t,k)$, and $w_i(t,k)$ 
to $c_{j^*}^*(t,k)$, $R_{j^*}(t,k)$,
and $w_{j^*}(t,k)$, respectively.

\subsection{Transformation region and comparative 
structural advantage}

The switching equilibrium tells us how an optimal economic
structure is determined for the given $(t,k)$. We now consider
the issue of finding the set of capital intensities to 
support structural transformation. Specifically, let 
$\mathscr{E}_i \in \mathcal{I}_{\mathscr{E}}$
be the economic structure prior to time $t$. 
We define the transformation region of capital 
intensities from $\mathscr{E}_i$ as follows.
\begin{equation}\label{sec5.transregion.equ1}
{\cal S}_i(t) = \Big\{ k \in (0, \infty) \ : \ 
\max_{j\neq i} V_j(t,k)- V_i(t,k) = 0 \Big\}.
\end{equation}

\noi ${\cal S}_i(t)$ is a closed subset of $(0, \infty)$ 
and represents a set of capital intensities with which 
the social planner should transform the economic
structure away from $\mathscr{E}_i$. Consider an economic 
structure $\mathscr{E}_j$, which is different from 
$\mathscr{E}_i$, and define
\begin{equation}\label{sec5.transregion.equ2}
{\cal S}_{i,j}(t)= \big\{ k \in {\cal S}_i(t) 
\ : \  V_j(t,k)=V_i(t,k) \big\}.
\end{equation}

\noi It is easy to see that
\begin{equation}\label{sec5.transregion.equ3}
{\cal S}_i(t) =  \bigcup_{ j\in \mathbbm{I}, j\neq i}
{\cal S}_{i,j}(t).
\end{equation}

\noi We also define ${\cal N}_i(t)$ as the complement set of
${\cal S}_i(t)$ in $(0, \infty)$, which is the so-called 
{\it continuation region} or the {\it no-transformation
region} associated with economic structure $\mathscr{E}_i$.
\begin{equation}\label{sec5.transregion.equ4}
{\cal N}_i(t) = \Big\{ k \in (0, \infty) \ : \ 
\sup_{j \neq i} V_j(t,k) < V_i(t,k) \Big\}.
\end{equation}

\noi ${\cal N}_i(t)$ is an open set and represents a 
collection of capital intensities with which economic 
structure $\mathscr{E}_i$ is optimal. In this open domain, 
the value function $V_i(t,k)$ is continuous differentiable 
and satisfies equation \eqref{sec5.stat.equil.equ1}. 
Then, by definition, 
\begin{equation}\label{sec5.transregion.equ5}
{\cal S}_i(t) \cup {\cal N}_i(t) = 
\Big[ \bigcup_{ j\in \mathbbm{I}, j\neq i}
{\cal S}_{i,j}(t) \Big] \cup {\cal N}_i(t)= (0, \infty).
\end{equation}

The definition of transformation regions suggests 
a measure for comparing the comparative advantage of 
an economic structure over another. 
Assume that the current economic structure is 
$\mathscr{E}_i$ and consider economic structures 
$\mathscr{E}_j$ and $\mathscr{E}_l$ $(j \neq l, j \neq i)$.
Let
\begin{equation}\label{comp.struct.adv.equ3}
{\cal H}{j, l}(t,k) := V_j(t,k) - V_l(t,k), \quad l \neq j, j \neq i.
\end{equation}

\noi ${\cal H}{j, l}(t,k)$ measures the comparative 
structural advantage of $\mathscr{E}_j$ over $\mathscr{E}_l$. 
Therefore, we say that economic structure 
$\mathscr{E}_j$ dominates economic structure $\mathscr{E}_l$ 
if ${\cal H}_i^{j, l}(t,k)>0$.

\section{Complex economic structures and stagewise development}

The model of catching-up economies in Xing (2023a) and the
above extended model can be further extended to 
incorporate more complicated structures. The mathematical
underpinnings also provides us a general framework to ``paste 
together" models of economic development and growth at 
different development stages via transformation of 
economic structures, so that various economic issues and 
different economic ideas can be discussed on the same platform. 
Instead of diminishing return to scale production function, 
Xing (2023b) studied the optimal growth of AK catching-up 
economies with human capital and explained the stagwise 
balanced and unbalanced feature of the competitive equilibrium
path. Xing (2023c) studied the 
optimal growth of catching-up economies with government
spending and explained why the standard form of Ricardian 
doctrine doesn't hold and needs to be modified. 
Xing (2023d) characterizes structural transformation and
structural switching in catching-up economies and explains
the distinction of structural transformation between frontier
and catching-up economies.  

Production function switching in catching-up economies 
(Xing, 2023abcd) can be explained as switching of a 
production structure. In this section, we further explain 
switching of economic structure by discussing how to
incorporate complex economic substructures, including composite 
production structures, choice of exogenous technology, 
switching between 
technology adoption and R\&D, and institutional structures.

\subsection{Intermediate goods and hierarchical 
production structures}

We first consider an economy in which intermediate goods and 
a unique final good are produced and the final good uses 
intermediate goods as inputs. Using a variant of the 
non-balanced growth model (Acemoglu and Guerrieri, 2008) 
as building blocks, the economy experiences stage-wise 
economic growth with EST.

\subsubsection*{Intermediate goods and technology}

The economy is similar to that in section 3, except 
intermediate goods are also produced. The final good is 
produced competitively by combining $m$ intermediate goods 
with elasticity of substitution $\epsilon \in [0, \infty)$, 
that is,
\begin{equation}\label{nonbal.prod.fun.equ1}
Y(t) = F[Y_1(t), \dots, Y_m(t); \omega] 
= \Big ( \sum_{j=1}^m w_j Y_j(t; \omega)^{\frac{\epsilon-1}{\epsilon}}
\Big)^{\frac{\epsilon}{\epsilon-1}},
\end{equation}

\noi where $w_1 > 0, \dots, w_m > 0$, $w_1 + \dots + w_m = 1$, 
and $\omega$ represents the vector of parameters in the
production functions and will be specified later. 
The intermediate goods $Y_1, \dots, Y_m$ are produced
competitively with production functions
\begin{equation}\label{nonbal.prod.fun.equ2}
Y_j(t; \omega) = F_i(K_j, L_j, A_j; \omega)=A_j(t) L_j(t)^{1-\alpha_j} K_j(t)^{\alpha_j},
\end{equation}

\noi where $A_j(t)$ is the technology of the $j$th 
intermediate good at time $t$. Technological progress
in all sectors is exogenous and takes the form
\begin{equation}\label{nonbal.tech.equ1}
\dot{A}_j(t) = \vartheta_j A_j(t), \qquad \vartheta_j > 0,
j=1, \dots, m.
\end{equation}

\noi For convenience, we assume that 
$\alpha_1 \le \cdots \le \alpha_m$, that is, the sectors
with larger $\alpha$'s are more capital intensive. 
Capital
and labor market clearing requires that at each time 
\begin{equation}\label{nonbal.prod.fun.equ3}
K_1(t) + \dots + K_m(t) \le K(t), \qquad
L_1(t) + \dots + L_m(t) \le L(t),
\end{equation}

\noi where $K$ denotes the aggregate capital stock and
$L$ is total population. $L_j(t)$ and $K_j(t)$
$(j=1, \dots, m)$ are nonnegative. Labor $L(t)$ is supplied 
inelastically and follows the process $L(t)=L(0) \exp(\pi t) $.

Let $P_j(t;\omega)$ be the price of $Y_j$ good 
$j$ $(j=1, \dots, m)$ at time $t$ for the given industrial 
structure $\omega$. We normalize the price of the final
good, $P$, to one at all points, so that
\begin{equation}\label{price.norm.equ1}
1\equiv P(t;\omega) = \Big( \sum_{j=1}^m w_{j}^\epsilon 
P_j(t; \omega)^{1-\epsilon}
\Big)^{1/(1-\epsilon)}.
\end{equation}

\noi Denote the rental price of capital, wage rate, and 
interest rate by $R(t; \omega)$, $w(t; \omega)$, and 
$r(t; \omega)$, respectively.

\subsubsection*{Production structures}

We now represent the production structure of the economy, 
using the notation introduced in section 3. Since $m$ 
intermediate goods and a final good are produced,
production and its structure can be expressed as 
$(\mathcal{Y}, \mathscr{Y}(\omega))$, where 
$\mathcal{Y} =\{ (Y(t), Y_1(t), \dots, Y_m(t)) \}$
and $\mathscr{Y}(\omega) =\{ F, F_1, \dots, F_m \}$, 
and $\omega=(w_1, \dots, w_m, \alpha_1, \dots, \alpha_m)$
is an element of the set $\Omega_{\mathscr{Y}} 
=\{ \omega \ | \ w_1,
\dots,w_m \ge 0, \sum_{j=1}^m w_j=1, 0\le \alpha_1 \le 
\dots < \alpha_m < 1 \}$. Here, $\mathscr{Y}(\omega)$
describes the composition and organization of industrial
structures, producing 
the final good and intermediate goods. Note that 
$w_j=0$ implies that intermediate goods $j$ are not
produced in the economy. Hence, the change of $w_j$ from 
$w_j=0$ to $w_j>0$ suggests that the composition of 
intermediate goods in the economy is changed. 

Exgoenous technology and its structure of production 
are expressed as $(\mathcal{A}, \mathscr{A})$, where 
$\mathcal{A}=\{ (A_1(t), \dots, A_m(t) \}$, 
$\mathscr{A}(\widetilde{\vartheta})
=\{ (A_{1,\vartheta_1}(\cdot), \dots, 
A_{m,\vartheta_m}(\cdot)) | A_j(\cdot) \mbox{ are given 
by \eqref{nonbal.tech.equ1}}\}$, and 
$\widetilde{\vartheta}=(\vartheta_1, \dots, \vartheta_m) 
\in \Omega_{\mathscr{A}}
=\{ \widetilde{\vartheta} | \vartheta_j >0, j=1,\dots, m \}$. 
Since at most $m$ intermediate goods are produced,  
labor allocation in production and its structure are
given by $(\mathcal{L}, \mathscr{L})$, where
$\mathcal{L}=\{ (L(t), L_1(t), \dots, L_m(t) \}$ and
$\mathscr{L}=\{ (L(\cdot), L_1(\cdot), \dots, L_m(\cdot)) |
L_1(\cdot) + \dots + L_m(\cdot) \le L(\cdot), 
L(\cdot) \mbox{ satisfies } \eqref{sec3.popu1} \}$.
Similarly, the allocation of capital stock and its 
structure are expressed as $(\mathcal{K}, \mathscr{K})$,
where $\mathcal{K}=\{(K(t), K_1(t), \dots, K_m(t)) | 
K_1(t)+\dots+K_m(t)\le K(t) \}$ and
$\mathscr{K}=\{ (K(\cdot), K_1(\cdot), \dots, K_m(t)) | 
K_1(\cdot)+\dots+K_m(\cdot)\le K(\cdot), K(\cdot)$
satisfies the budget constraint 
\eqref{sec3.aggre.capital.equ1}$\}$.
Goods and factor prices and their structure can be 
represented as $(\mathcal{P}, \mathscr{P})$, where 
$\mathcal{P}=\{(R(t), w(t), P_1(t), \dots, P_m(t), P(t) ) \}$
and $\mathscr{P}=\{$factor prices are determined by firms
to maximize their profit, and $P_1(t), \dots, P_m(t),
P(t)$ satisfy \eqref{price.norm.equ1}$\}$.

Then with $(\mathcal{H}, \mathscr{H})$, 
$(\mathcal{F}, \mathscr{F})$, $(\mathcal{M}, \mathscr{M})$,
$(\mathcal{C}, \mathscr{C})$, and $(\mathcal{U}, \mathscr{U})$
defined as those in section 3, agents' behavior
and economic activities and the economic structure in 
the economy can still be expressed as $(\mathcal{E}, 
\mathscr{E})$. To highlight the structures of production
and exogenous technological progress, which are described
by $\omega \in \Omega_{\mathscr{Y}}$ and 
$\widetilde{\vartheta} \in \Omega_{\mathscr{A}}$, 
respectively, the economic structure of the economy may
be expressed as $\mathscr{E}(\omega, \widetilde{\vartheta})$.

\subsubsection*{Social planner's objective and 
competitive equilibrium}

Suppose the social planner's information set
of economic structures is $\mathcal{I}_{\mathscr{Y}
\times \mathscr{A}}=\{ \mathscr{E}_i |\mathscr{E}_i=
\mathscr{E}(\omega_i, \widetilde{\vartheta}_i), i
\in \mathbbm{I} \}$.
The social planner can choose an economic structure for 
the economy and allocate resources under the chosen 
economic structure.
Assume that an economic structure $\mathscr{E}_i \in 
\mathcal{I}_{\mathscr{Y}\times \mathscr{A}} $
(or equivalently, $i\in \mathbbm{I}$) is chosen 
for the economy. The aggregate resource constraint, 
which is equivalent to the budget constraint of the 
representative household, has the same form as 
\eqref{sec3.aggre.capital.equ1}. Define the shares 
of capital and labor allocated to industry $i$ as 
$\xi_j(t) \equiv \frac{K_j(t)}{K(t)}$, 
and $\lambda_j(t) \equiv \frac{L_j(t)}{L(t)}.$
Then output per capita at time $t$ can be
expressed as
\begin{equation}\label{nonbal.capital.inten.equ1}
y(t; \omega) \equiv 
f(t,k(t); \omega_i) \equiv  \Big\{ 
\sum_{j=1}^m w_{ij} \big[ k(t)^{\alpha_{ij}} A_j(t)
\xi_j(t)^{\alpha_{ij}} \lambda_j(t)^{1-\alpha_{ij}} 
\big]^{\frac{\epsilon-1}{ \epsilon}} \Big\}^{ 
\frac{\epsilon}{\epsilon-1}} .
\end{equation}

At each time $t$, the social planner in the economy must
optimally choose an economic structure $\theta(t)=
\mathscr{E}_i \in \mathcal{I}_{\mathscr{Y}\times \mathscr{A}}$, 
consumption $c(t)$, and allocations of capital and labor 
$(\xi_j(t), \lambda_j(t))$. The aggregate resource constraint 
per capita is 
\begin{equation}\label{sec6.per.capital.equ3}
\left\{ \begin{array}{ll}
\dot{k}(t) = f_{\kappa_n}(t, k(t); \omega_i) - 
(\delta+\pi) k(t) - c(t), & 
\tau_n \le t < \tau_{n+1}, \\
k(\tau_n-) = k(\tau_n), &
t=\tau_{n+1}, n=1, 2, \dots
\end{array} \right.
\end{equation}

\noi The representative household's total utility, starting at 
time $t_0$, has the same form as \eqref{sec3.totalutil.equ1}
and is given as
\begin{equation}\label{sec6.totalutil.equ1}
J_i(t_0, k; \{c(t), \xi \} )= 
\int_{t_0}^\infty e^{-(\rho-\pi) t} u(c(t)) dt
\end{equation}

\noi The social planner's objective is to solve the maximization 
problem 
\begin{equation}\label{sec6.totalutil.equ2}
\begin{aligned}
& V_i(t_0, k) = \max_{ \{ c(t), \xi\} }  
J_i(t_0, k; \{c(t), \xi \} ) \\
& \qquad
\qquad \mbox{subject to \eqref{sec6.per.capital.equ3}} 
\mbox{  and } k(t_0) = k\in \mathbbm{R}^+, 
\theta(t_0)=\mathscr{E}_i \in 
\mathcal{I}_{\mathscr{Y}\times \mathscr{A}}.
\end{aligned}
\end{equation}

Given the above specification, the {\it competitive 
equilibrium} of the economy consists of paths for factor 
and intermediate goods prices $\{ r(t; \theta(t)), 
w(t; \theta(t)), p_j(t; \theta(t))\}$, employment and 
capital allocation $\{\xi_j(t; \theta(t)),
\lambda_j(t; \theta(t))\}$, consumption and savings 
decisions $\{ c(t), \dot{K}(t) \}$, and economic 
structures $\{ \theta(t) \}$ such that the utility of 
the representative household is maximized, firms 
maximize profits, and markets clear. 
To characterize the competitive equilibrium mathematically, 
we first express the maximization problem in the form of the 
combined optimal control and optimal switching problem in 
section 4.1 in which 
$$
\mu_i(t,k,c) = f(t,k(t); \omega_i) - (\delta+\pi)k(t)-c(t), 
\qquad \tau_n \le t  < \tau_{n+1}, 
$$

\noi and the control space 
$$
\begin{aligned}
\mathfrak{U} & \equiv \{ (c(t), \xi_1(t), \dots, \xi_I(t), 
\lambda_1(t), \dots, \lambda_I(t)) \ | \ c(t): [0, \infty) 
\rightarrow U, \xi_i(t): [0, \infty) \rightarrow [0,1], \\
& \hspace{2cm} \lambda_i(t): [0, \infty) \rightarrow [0,1] 
\mbox{ are Lebesgue measurable functions} \}.
\end{aligned}
$$

\noi Using the method in Xing (2021), we have
the following:

\begin{proposition}\label{sec6.nonbal.prop1}
{\it For each $i\in \mathbbm{I}$, $V_i(t,k)$ defined by 
\eqref{sec3.totalutil.equ2} is a viscosity solution to
\begin{equation}\label{sec6.hjbqvi.def1}
\begin{aligned}
& \max\Big\{ \sup_{c(t), \lambda_j(t), \xi_j(t)} \Big[ 
\frac{\partial V_i}{\partial t}(t,k) + \big[ f_i(t,k)-
(\delta+\pi)k -c \big] \frac{\partial V_i}{\partial k}(t,k) 
+ e^{-(\rho-\pi)t} u(c) \Big], \\
& \hspace{100pt} \max_{j\neq i} V_j(t,k)
-V_i(t,k) \Big\} = 0,
\end{aligned}
\end{equation}

\noi and such solutions are unique. In particular, 
the static and dynamic equilibria of the economy 
are characterized by the supremum in \eqref{sec6.hjbqvi.def1}, 
and the switching equilibrium is characterized by the 
maximum over $j\neq i$ in \eqref{sec6.hjbqvi.def1}. }
\end{proposition}

Our model extends the two-sector nonbalanced
growth model in Acemoglu and Guerrieri (2008) to the case
of an $m$-sector stagewise nonbalanced growth model with
EST. The static and dynamic equilibria in 
\eqref{sec6.hjbqvi.def1} can be analyzed similarly as 
in Acemoglu and Guerrieri (2008).

\subsubsection*{Economic implication}

The extended EST model of stagewise growth characterizes 
an economy's endogenous transformation among various 
production structures and has different variants if 
more detailed assumptions about the economy can be provided.
We now briefly discuss several EST scenarios of the model.

First, consider a problem of economic development in 
which the economy is at the early development stage and 
produces only a few intermediate goods. For example, assume 
the economy produces two intermediate goods, that is, 
$\omega=(w_1, w_2, 0, \dots, 0)$, with 
$0<\alpha_1 < \alpha_2<1$, so the production structure is
$\mathscr{Y}(\omega)$. This is the case of non-balanced
growth studied by Acemoglu and Guerrieri (2008).
When the number of intermediate goods in the world is 
greater than two, that is, $I > 2$, development of
the economy can be achieved by transforming the production
structures from two to three intermediate products. Suppose 
the economy transforms from $\mathscr{Y}(\omega)$ to 
$\mathscr{Y}(\omega')$, where $\omega'=(w_1', w_2', 
w_3', 0, \dots, 0)$ and $0<\alpha_1<\alpha_2<\alpha_3<1$.
Usually two issues are involved in the transformation:
one is when to start producing the third intermediate good
and the other is which intermediate goods should be chosen.
Using the extended EST model and competitive equilibrium
argument here, these two issues can be solved.

Second, consider a problem of economic transition in which
the country gives priority to development of capital intensive
heavy industries when capital in the country is scarce. 
Since the optimal economic structure is dependent on the 
level of capital intensity, the process of economic transition
is equivalent to that of structural transformation in 
which more economic resources need to be allocated to
less capital intensive industries, and the path of 
optimal economic structures characterizes the process
of transition of production structures. Such transition
process also involves transformation of institutional 
structures, which will be discussed briefly in section 
6.5. 

Third, the extended model here also sheds a light on 
poverty trap and/or middle-income trap problems. 
Most poverty trap models in the literature argue the
existence of poor and nonpoor equilibria and discuss
the possibilities of moving from the poor to the
non-poor equilibria. They seldom study the poverty
trap problem by equilibrium arguments on transformation 
of economic structures. The extended EST model and its
competitive equilibirum can be used to characterize
the poor and nonpoor (static and dynamic) equilibria
under different economic structures and the transition
process of economic structures (or structural equilibirum).
In addition to the poverty trap problem, the middle-income 
trap problem can be similarly discussed when the 
economy in the EST model is further extended to 
incorporate countries' trade structures in open 
economies.

\subsection{Structures of technological progress}

Economic history shows that technological progress in 
countries on the world's production possibility frontier
is usually achieved via R\&D, whereas technological 
progress in countries inside the frontier
may be attained via technology adoption and/or R\&D. As 
previous EST models assume exogenous technological progress,
we extend the EST to endogenize structural transformation
of different types of technological progress and discuss
its economic implications. 

\subsubsection*{Production with variety and quality ladders}

The literature usually considers two types of 
technological change, namely process and product 
innovation, or the introduction of a new product and 
innovations to reduce the costs of production of existing
products. Two commonly used canonical models for these
changes are product-variety models (Romer, 1990) and 
Schumpeterian models (Aghion and Howitt, 1992). 
For convenience, we extend production function 
\eqref{nonbal.prod.fun.equ1} and 
assume that the unique final good is produced
competitively by combining the output of 
$m(t)$ sectors with elasticities of substitution
$\epsilon \in [0, \infty)$, that is, 
\begin{equation}\label{tech.prod.fun.equ1}
Y(t; \omega) = F[Y_1(t), \dots, Y_{m(t)}(t)] 
= \Big ( \int_0^{m(t)} w(j,t) q(j, t)
Y(j, t)^{\frac{\epsilon-1}{\epsilon}}
dj \Big)^{\frac{\epsilon}{\epsilon-1}},
\end{equation}

\noi where the weights $w(j,t)$ satisfy $w(j,t) \ge 0$
for all $t$ and $j$ and $\int_0^{m(t)} w(j,t)dj =1$,
$m(t)$ denotes the number of varieties of inputs
at time $t$, $q(j, t)$ represents the ``quality
ladder" for machine type $j$, and $\omega$ is a vector
of parameters related to the production functions and will be
specified later. The intermediate goods of machine type $j$,
$Y(j,t)$, are produced by the production function
\begin{equation}\label{tech.prod.fun.equ2}
Y(j, t) :=F(j,t)= K(j,t)^{\alpha(j,t)}L(j,t)^{1-\alpha(j,t)}.
\end{equation}

\noi The technology level of machine type $j$ is represented 
by the quality ladder $q(j,t)$; hence, there is no need to 
keep the term $A(j,t)$ in \eqref{tech.prod.fun.equ2}. 
Let $\xi(j,t; \omega) = K(j,t; \omega)/K(t)$ and 
$\lambda(j,t; \omega) = L(j,t; \omega)/ L(t)$.  
The output per capita at time $t$ is 
\begin{equation}\label{tech.capital.inten.equ1}
f(t,k(t); \omega) =  \Big\{ \int_0^{m(t)}
w(j,t) q(j,t) \big[ k(t)^{\alpha(j,t)} 
\xi_j(t)^{\alpha(j,t)} \lambda_j(t)^{1-\alpha(j,t)} 
\big]^{\frac{\epsilon-1}{ \epsilon}} \Big\}^{ 
\frac{\epsilon}{\epsilon-1}} .
\end{equation}

\noi Capital and labor market clearing requires that 
\begin{equation}\label{tech.prod.fun.equ3}
\int_0^{m(t)} K(j,t)dj \le K(t), \qquad
\int_0^{m(t)} L(j,t)dj \le L(t).
\end{equation}

\noi Then production and its structure of the economy
can be expressed as $(\mathcal{Y}, \mathscr{Y}(\omega))$, 
where $\mathcal{Y}=\{ Y(t), \{ Y(j,t) \}_{0\le j\le m(t)} 
\}$ describes the output levels of the intermediate and 
final goods, $\mathscr{Y}(\omega)=\{F(\cdot), 
\{F(j,\cdot)\}_{0 \le j \le m(t)} \}$ represents the 
composition and organization of all production, and $\omega=
(\{w(j,\cdot)$, $\alpha(j,\cdot) \}_{0\le j\le m(t)})$
is an element of the set $\Omega_{\mathscr{Y}}=
\{ \omega | w(j,t)\ge 0$, $\int_0^{m(t)} w(j,t)dj=1$,
$0< \alpha(j,t) < 1$ for all $j\}$.

\subsubsection*{Endogenous technological progress}

We first consider endogenous technological progress
via technology adoption. Suppose $\widetilde{m}(t)$
and $\widetilde{q}(j, t)$ are the number of machines and
quality ladder for machine type $j$ on the world's 
production possibility frontier at time $t$, respectively.
By definition, $\widetilde{m}(t)$ and $\widetilde{q}(j, t)$ 
are exogenously determined and monotonically nondecreasing 
over $t$. 
The economy in our study has $m(t-)$ machines and machine 
type $j$ is on the quality ladder $q(j, t-)$ prior to 
time $t$. Obviously, $m(t-) \le  \widetilde{m}(t)$ and
$q(j, t-) \le \widetilde{q}(j, t)$. 

At time $t$, the social planner of the economy decides 
whether the number of machine types and the quality of each 
machine type should be improved via technology adoption. 
\footnote{Recall that the social planner is defined as 
the social elite of an economy, which can collect
information and make decision on production and other 
economic activities.}
If yes, the number of machine types $m(t)$ and 
the quality ladder $q(j,t)$ of machine type $j$ at time
$t$ should be chosen from the sets $(m(t-), \widetilde{m}(t)]$
and $(q(j,t-), \widetilde{q}(j,t)]$, respectively. 
Then the level of technology is given by
$$
\mathcal{A}=\{ (m(t), \{q(j,t)\}_{0\le j \le m(t)}), 
\mbox{ where } m(t)\in (m(t-), \widetilde{m}(t)]
\mbox{ and } q(j,t) \in (q(j,t-), \widetilde{q}(j,t)]\},
$$

\noi and the technological structure in production is 
expressed as $\mathscr{A}_{\mathrm{adopt}}(\epsilon)=\{
\epsilon=(\epsilon_m, \{ \epsilon_j \}_{0\le j\le \epsilon_m}
)$, $\epsilon_m = m(\cdot)$, $\epsilon_j=q(j, \cdot)\}$, 
which is an element of the information set 
$\mathcal{I}_{\mathscr{A}, \mathrm{adopt}}
=\{ \mathscr{A}_{\mathrm{adopt}}(\epsilon) | \epsilon \in 
\Omega_{\mathscr{A}, \mathrm{adopt}} \}$ and
$\Omega_{\mathscr{A}, \mathrm{adopt}}
=\{ (\epsilon_m, \{\epsilon_j\}_{0\le j \le \epsilon_m})
| \epsilon_m \in (m(t-), \widetilde{m}(t)], 
\epsilon_j \in (q(t,j), \widetilde{q}(t,j)] \}.$
A mathematical concern here is whether such a decision process
leads to continuous variation of $m(t)$ and $q(j,t)$,
which indicates that the technological structure violates
the durationality attribute. This concern only arises in 
theoretical analysis, since the improvement from $m(t-)$ 
to $m(t)$ and/or from $q(j,t-)$ to $q(j,t)$ in reality involves
some cost, which prevents the social planner from choosing
$m(t)$ and $q(j,t)$ continuously. In theoretical analysis,
the diminishing marginal returns of production functions 
ensure durationality and transformality. Hence, 
the functionals $\mathscr{A}_{\mathrm{adopt}}$, chosen 
optimally by the social planner, are still piecewise 
constant. Then given production $(\mathcal{Y}, 
\mathscr{Y}(\omega))$ and technology $(\mathcal{A}, 
\mathscr{A}_{\mathrm{adopt}})$, the resource constraint
of the economy at time $t$ is 
\begin{equation}\label{tech.prod.adopt.equ1}
\dot{K}(t) = Y(t; \omega) - \delta K(t) - C(t).
\end{equation}

\noi Accordingly, the allocation of capital stock and its 
structure are expressed as $(\mathcal{K}, 
\mathscr{K}_{\mathrm{adopt}}(\omega, \epsilon))$, where 
$\mathcal{K}$ is defined similarly as in section 6.1 and 
$\mathscr{K}_{\mathrm{adopt}}(\omega, \epsilon)=\{ K(j,\cdot) | 
\int_0^{m(t)} K(j,\cdot)dj \le K(\cdot)$, $K(\cdot)$
satisfies \eqref{tech.prod.adopt.equ1}$\}$.

In addition to technology adoption, the social planner may also
decide to improve the technology level via R\&D. To explain
how to describe the structure of technological progress via 
R\&D, we 
suppose that $Z(t)$ is expenditure on R\&D at
time $t$, and the number of machine types $m(t)$ 
and the quality of the $j$th machine type $q(j,t)$
at time $t$ satisfy the following 
\begin{equation}\label{rd.prod.rnd.equ2}
\dot{m}(t)=\iota_m Z(t), \qquad
\lim_{\Delta t\rightarrow 0} 
\frac{q(j,t+\Delta t)-q(j,t)}{\Delta t}
= \iota_j Z(t) \quad \mbox{for all }  j
\mbox{ and } t,
\end{equation}

\noi where $\iota_m$ and $\iota_j$ are nonnegative parameters 
and satisfy the constraint $\iota_m + \int_0^{m(t)} 
\iota_j dj = 1$. Then the technology level of production 
is $\mathcal{A}=\{ (m(t), \{q(j,t)\}_{0\le j \le  m(t)}) 
| m(t)$ and $q(j,t)$ satisfy \eqref{rd.prod.rnd.equ2}$\}$.
To represent the technological structure, let 
$\iota=(\iota_m, \{\iota_j\})$ and $\Omega_{\mathscr{A},
\mathrm{r\&d}} = \{ \iota | \iota_m\ge 0, \iota_j \ge 0 
\mbox{ for all }j, \iota_m + \int_0^{m(t)} \iota_j dj = 1 \}$. 
Then the structure of technology can be expressed as
$\mathscr{A}_{\mathrm{r\&d}}(\iota)$, which is an element
of the set of functionals $\mathcal{I}_{\mathscr{A}, 
\mathrm{r\&d}}=\{ (m(\cdot), \{q(j, \cdot) \}_{
0\le j\le m(\cdot)}) | \iota \in \Omega_{\mathscr{A}, 
\mathrm{r\&d}}$, $m(\cdot)$ and $q(j,\cdot)$ satisfy 
\eqref{rd.prod.rnd.equ2}$\}$. Provided production 
$(\mathcal{Y}, \mathscr{Y}(\omega))$ and technology 
$(\mathcal{A}, \mathscr{A}_{\mathrm{r\&d}}(\iota))$, the 
resource constraint of the economy at time $t$ is 
\begin{equation}\label{rd.prod.rnd.equ1}
\dot{K}(s) = Y(t; \omega) - \delta K(t) - 
C(t) - Z(t).
\end{equation}

\noi Consequently, the allocation of capital stock and its 
structure are expressed as $(\mathcal{K},
\mathscr{K}_{\mathrm{r\&d}}(\omega, \iota))$, 
where $\mathcal{K}$ is defined similarly as in section 6.1 and 
$\mathscr{K}_{\mathrm{r\&d}}(\omega, \iota)=\{ K(j,\cdot) | 
\int_0^{m(t)} K(j,\cdot)dj \le K(\cdot)$, $K(\cdot)$
satisfies \eqref{rd.prod.rnd.equ1}$\}$.

Therefore, the total information set of technology and its 
structure for the social planner is $\mathcal{I}_{\mathscr{A}}
:= \mathcal{I}_{\mathscr{A}, \mathrm{adopt}} \cup 
\mathcal{I}_{\mathscr{A}, \mathrm{r\&d}}$, and the
corresponding total information set of capital allocation
and its structure is $\mathcal{I}_{\mathscr{K}}
:= \mathcal{I}_{\mathscr{K}, \mathrm{adopt}} \cup 
\mathcal{I}_{\mathscr{K}, \mathrm{r\&d}}$. Once the
social planner chooses structures of technology and 
capital allocation, technology level $\mathcal{A}$
and capital alloation $\mathcal{K}$
can be determined by corresponding mechanisms. 

Other structures $(\mathcal{H}, \mathscr{H})$, 
$(\mathcal{F}, \mathscr{F})$, $(\mathcal{M}, \mathscr{M})$,
$(\mathcal{L}, \mathscr{L})$,
$(\mathcal{P}, \mathscr{P})$,
$(\mathcal{C}, \mathscr{C})$, and 
$(\mathcal{U}, \mathscr{U})$ can be defined similarly
with the necessary modifications as in section 6.1. Then 
we may still use $(\mathcal{E}, \mathscr{E})$ to 
represent agents' behavior and economic activities 
and their structures in the economy. To highlight 
production structure $\omega \in \Omega_{\mathscr{Y}}$ 
and technological structure 
$\epsilon \in \Omega_{\mathscr{A}, \mathrm{adopt}}$
or $\iota \in \Omega_{\mathscr{A}, \mathrm{r\&d}}$, 
one may express the economic structure of the economy
as $\mathscr{E}(\omega, \epsilon)$ or 
$\mathscr{E}(\omega, \iota)$. 

\subsubsection*{Social planner's maximization problem}

Suppose the social planner's information set
of economic structures is $\mathcal{I}_{\mathscr{Y}
\times \mathscr{K} \times \mathscr{A}}=\{ \mathscr{E}_i 
|\mathscr{E}_i= \mathscr{E}(\omega_i, 
\widetilde{\vartheta}_i), \widetilde{\vartheta}_i \in 
\Omega_{\mathscr{A}, \mathrm{adopt}} \cup
\Omega_{\mathscr{A}, \mathrm{r\&d}}, i\in \mathbbm{I} \}$.
The social planner can choose an economic structure for 
the economy and allocate resources under the given 
economic structure. When $\mathscr{E}(\omega_i, 
\widetilde{\vartheta}_i)$ $(\widetilde{\vartheta}_i \in 
\Omega_{\mathscr{A}, \mathrm{adopt}})$ is chosen, 
equation \eqref{tech.prod.adopt.equ1} implies that
the resource constraint per capita is
\begin{equation}\label{tech.prod.adopt.equ2}
\dot{k}(t) = f(t,k(t); \omega_i) - (\delta+\pi) k(t) - c(t),
\end{equation}

\noi and when $\mathscr{E}(\omega_i, 
\widetilde{\vartheta}_i)$ $(\widetilde{\vartheta}_i \in 
\Omega_{\mathscr{A}, \mathrm{r\&d}})$ is chosen, 
equation \eqref{rd.prod.rnd.equ1} implies that the
resource constraint per capita is
\begin{equation}\label{rd.prod.rnd.equ3}
\dot{k}(t) = f(t,k(t); \omega) - (\delta+\pi) k(t) - 
c(t) -z(t).
\end{equation}

\noi Suppose the economy starts with economic structure 
$\mathscr{E}(\omega_i, \widetilde{\vartheta}_i)$ at time 
$t_0$, Then the social planner's total utility 
starting at time $t_0$ is given by
$$
J_i(t_0, k; \{c(t), \xi \} )= 
\int_{t_0}^\infty e^{-(\rho-\pi) t} u(c(t)) dt. 
$$

\noi The social planner's objective is to solve the maximization 
problem 
\begin{equation}\label{tech.totalutil.equ2}
\begin{aligned}
& V_i(t_0, k) = \max_{ \{ c(t), \xi\} }  
J_i(t_0, k; \{c(t), \xi \} ) \qquad \mbox{subject to 
\eqref{tech.prod.adopt.equ2}} \mbox{ or 
\eqref{rd.prod.rnd.equ3}}\\
& \qquad \qquad \qquad
\mbox{  and } k(t_0) = k\in \mathbbm{R}^+, 
\theta(t_0)=\mathscr{E}_i \in 
\mathcal{I}_{\mathscr{Y}\times \mathscr{K}\times \mathscr{A}}.
\end{aligned}
\end{equation}

Given the above specification, the competitive equilibrium
of the economy consists of paths for factor and 
intermediate goods prices, employment and capital 
allocation $\{\xi_j(t;\theta(t)), \lambda_j(t; \theta(t))\}$, 
consumption and savings decisions $\{ c(t), \dot{K}(t) \}$, 
and structures of production, capital allocation, and
technological progress $\{ \theta(t)\}$ such that the 
utility of the representative household is maximized, 
firms maximize profits, and markets clear. 
Using the method in Xing (2021), we can show the following.

\begin{proposition}\label{sec6.rd.prop1}
{\it For each $\mathscr{E}_i= \mathscr{E}(\omega_i, 
\widetilde{\vartheta}_i) \in \mathcal{I}_{\mathscr{Y}
\times \mathscr{K} \times \mathscr{A}}$, let $V_i(t,k)$
be defined by \eqref{sec6.totalutil.equ2}. 
Then for $\widetilde{\vartheta}_i \in  \Omega_{\mathscr{A}, 
\mathrm{adopt}}$, $V_i(t,k)$ is a unique viscosity solution to
\begin{equation}\label{sec6.rd.hjbqvi.def1a}
\begin{aligned}
& \max\Big\{ \sup_{e(t)} \Big[ 
\frac{\partial V_i}{\partial t}(t,k) + \big[ f_i(t,k)-
(\delta+\pi)k -c \big] \frac{\partial V_i}{\partial k}(t,k) 
+ e^{-(\rho-\pi)t} u(c) \Big], \\
& \hspace{50pt} \max_{j\neq i} \big\{ 
V_j(t,k) \big| \widetilde{\vartheta}_j \in 
\Omega_{\mathscr{A}, \mathrm{r\&d}} \cup
\Omega_{\mathscr{A}, \mathrm{adopt}} \big\}
-V_i(t,k) \Big\} = 0,
\end{aligned}
\end{equation}

\noi where $e(t) = (c(t), \{ \xi(j,t), \lambda(j,t) \})$.
For $\widetilde{\vartheta}_i \in \Omega_{\mathscr{A}, 
\mathrm{r\&d}}$, $V_i(t,k)$ is a unique viscosity solution to
\begin{equation}\label{sec6.rd.hjbqvi.def1b}
\begin{aligned}
& \max\Big\{ \sup_{e(t) } \Big[ 
\frac{\partial V_i}{\partial t}(t,k) + \big[ f_i(t,k)-
(\delta+\pi)k -c-z \big] \frac{\partial V_i}{\partial k}(t,k) 
+ e^{-(\rho-\pi)t} u(c) \Big], \\
& \hspace{50pt} \max_{j\neq i} \big\{ 
V_j(t,k) \big| \widetilde{\vartheta}_j \in 
\Omega_{\mathscr{A}, \mathrm{r\&d}} \cup
\Omega_{\mathscr{A}, \mathrm{adopt}} \big\}
-V_i(t,k) \Big\} = 0,
\end{aligned}
\end{equation}

\noi where $e(t) = (c(t), z(t), \{ \xi(j,t), \lambda(j,t) \})$.}
\end{proposition}

\subsubsection*{Economic implications}

Proposition \ref{sec6.rd.prop1} integrates several 
scenarios of endogenous technological progress into 
a single framework and has interesting implication for
economic development and growth. We next briefly
discuss scenarios and implications of Proposition
\ref{sec6.rd.prop1}.

The first scenario is countries on the global production 
possibility frontier and their technological progress. 
Since these countries or economies are on the world 
production possibility frontier, the social planners of 
these economies will rule out the possibility of technological 
adoption and the information set of technological 
structure reduces from 
$\mathcal{I}_{\mathscr{A}, \mathrm{adopt}} \cup 
\mathcal{I}_{\mathscr{A}, \mathrm{r\&d}}$ to
$\mathcal{I}_{\mathscr{A}, \mathrm{r\&d}}$. 
Hence, the HJB-QVI system \eqref{sec6.rd.hjbqvi.def1b}
for the value function of the representative
household in Proposition \ref{sec6.rd.prop1} 
becomes that, for $\widetilde{\vartheta}_i 
\in \Omega_{\mathscr{A}, \mathrm{r\&d}}$, 
\begin{equation}\label{sec6.rd.hjbqvi.def2a}
\begin{aligned}
& \max\Big\{ \sup_{e(t) } \Big[ 
\frac{\partial V_i}{\partial t}(t,k) + \big[ f_i(t,k)-
(\delta+\pi)k -c-z \big] \frac{\partial V_i}{\partial k}(t,k) 
+ e^{-(\rho-\pi)t} u(c) \Big], \\
& \hspace{50pt} \max_{j\neq i} \big\{ 
V_j(t,k) \big| \widetilde{\vartheta}_j \in 
\Omega_{\mathscr{A}, \mathrm{r\&d}} \big\}
-V_i(t,k) \Big\} = 0.
\end{aligned}
\end{equation}

\noi This indicates that R\&D is the only way to achieve
technological progress for countries on the global production 
possibility frontier, and the structure of technological 
progress via R\&D is characterized by the rate $\iota
=(\iota_m, \{ \iota_j \}) \in \Omega_{\mathscr{A},
\mathrm{r\&d}}$. If a rate of R\&D is fixed with
$\iota$ and no transformation of the R\&D structure
is needed, then \eqref{sec6.rd.hjbqvi.def2a} 
further reduces to
\begin{equation}\label{sec6.rd.hjbqvi.def2b}
\sup_{e(t) } \Big[ 
\frac{\partial V_i}{\partial t}(t,k) + \big[ f_i(t,k)-
(\delta+\pi)k -c-z \big] \frac{\partial V_i}{\partial k}(t,k) 
+ e^{-(\rho-\pi)t} u(c) \Big]= 0,
\end{equation}

\noi which is in the form of technological progress
via R\&D in the neoclassical economy. 

The second scenario is developing countries that are 
inside the global production possibility frontier. 
Two situations might occur.
One is that the developing country is well connected with 
the world's economy, so that the information 
of the world's technology can be accessed by the economy
freely or with low cost. In such case, the social 
planner of the economy would naturally consider improving
the country's technology via adoption instead of R\&D, 
as inside-the-frontier indigenous innovation is not efficient.
Thus, the HJB-QVI system \eqref{sec6.rd.hjbqvi.def1b}
for the value function of the representative
household in Proposition \ref{sec6.rd.prop1} 
 becomes that, for $\widetilde{\vartheta}_i 
\in \Omega_{\mathscr{A}, \mathrm{adopt}}$, 
\begin{equation}\label{sec6.rd.hjbqvi.def3a}
\begin{aligned}
& \max\Big\{ \sup_{e(t)} \Big[ 
\frac{\partial V_i}{\partial t}(t,k) + \big[ f_i(t,k)-
(\delta+\pi)k -c \big] \frac{\partial V_i}{\partial k}(t,k) 
+ e^{-(\rho-\pi)t} u(c) \Big], \\
& \hspace{50pt} \max_{j\neq i} \big\{ 
V_j(t,k) \big| \widetilde{\vartheta}_j \in 
\Omega_{\mathscr{A}, \mathrm{adopt}} \big\}
-V_i(t,k) \Big\} = 0.
\end{aligned}
\end{equation}

\noi Some discussion starts from the perspective 
of the social planners in developed countries and focuses 
on technology diffusion. The social planners in these 
discussions play a passive role and 
accept the technology diffused from developed countries
without the process of choosing ``appropriate" technology
levels (or structures). Equation \eqref{sec6.rd.hjbqvi.def3a}
avoids this issue and highlights the active role of
social planners in developing countries in choosing
``appropriate" technological structures.
Another situation in the second scenario is that
the developing country is somehow isolated from the world 
economy, so that inside-the-frontier innovation 
is necessary. In such case, the value function
of the representative household in the isolated
economy is characterized by the HJB-QVI system 
\eqref{sec6.rd.hjbqvi.def2b}.

The third scenario is countries that are inside 
but near the global production possibility frontier.
As the technology level in those countries is near 
the global production possibility frontier, due to the 
fear of competition, the countries on the global 
production possibility frontier may embargo their 
technology knowhow. Therefore, 
R\&D must be carried out for further economic
development. Then the main issue for the social planners 
in these countries is when to switch from technology 
adoption to R\&D. To fix the idea, suppose the social 
planner's information set of technological structures 
is $\mathcal{I}_{\mathcal{A}} = \{ 
\mathscr{E}(\omega_1, \widetilde{\vartheta}_1), 
\mathscr{E}(\omega_2, \widetilde{\vartheta}_2)\}$ with 
$\widetilde{\vartheta}_1 \in \Omega_{\mathscr{A}, 
\mathrm{adopt}}$ and $\widetilde{\vartheta}_2 \in 
\Omega_{\mathscr{A}, \mathrm{r\&d}}$. Then Proposition
\ref{sec6.rd.prop1} implies that the value
functions $V_1(t,k)$ and $V_2(t,k)$ of the representative
household with corresponding initial economic structures
$\mathscr{E}(\omega_1, \widetilde{\vartheta}_1)$ and 
$\mathscr{E}(\omega_2, \widetilde{\vartheta}_2)$, 
respectively, and times of switching from the mode of
technology adoption to that of R\&D are characterized 
by the following HJB-QVI system
\begin{equation}\label{sec6.rd.hjbqvi.def4}
\begin{aligned}
\left\{ \begin{array}{l}
\max\Big\{ \displaystyle \sup_{e(t)} \Big[ 
\frac{\partial V_1}{\partial t}(t,k) + \big[ f_1(t,k)-
(\delta+\pi)k -c \big] \frac{\partial V_1}{\partial k}(t,k) 
+ e^{-(\rho-\pi)t} u(c) \Big], \\
\hspace{50pt} V_2(t,k) -V_1(t,k) \Big\} = 0, 
\qquad e(t)=(c(t), \{\xi(j,t), \lambda(j,t)\}); \\
\max\Big\{ \displaystyle \sup_{e(t) } \Big[ 
\frac{\partial V_2}{\partial t}(t,k) + \big[ f_2(t,k)-
(\delta+\pi)k -c-z \big] \frac{\partial V_i}{\partial k}(t,k) 
+ e^{-(\rho-\pi)t} u(c) \Big], \\
\hspace{50pt} V_1(t,k) -V_2(t,k) \Big\} = 0,
\qquad e(t)=(c(t), z(t), \{\xi(j,t), \lambda(j,t)\}).
\end{array} \right.
\end{aligned}
\end{equation}

\subsection{Infrastructure and economic institutions}

The discussion so far has focused on structures of production 
and consumption and their transformation in the process of 
economic development. In general, in addition to the process of 
industrial upgrading and technological progress, which involves 
households' and firms' decisions on the supply and demand
of factors of production, an economy's development
also involves the production of public goods and
infrastructure, which are supplied by governments or
require collective actions and cannot be internalized 
in the decisions of individual households or firms.
We now consider production structures of public goods 
and infrastructure. 

Infrastructure includes hard infrastructure and soft 
infrastructure. Hard infrastructure consists of the physical
infrastructure of highways, port facilities, airports,
telecommunication systems, electricity grids, and other
public utilities. Soft infrastructure consists of 
institutions, regulations, social capital, and other
social and economic arrangements. Most hard infrastructure
and almost all soft infrastructure is exogenously 
provided to individual firms in the form of public 
goods and cannot be internalized in their production 
decisions. To illustrate the idea, we consider an approach
to model infrastructure as public goods available 
to firms and a government's tax policy as a simplified 
economic institution. 
\footnote{Economic institutions have different meanings
in different contexts, and we refer to  them as taxes, the
security of property rights, contracting institutions,
and other economic arrangements. This is different from 
the political institutions discussed in the next subsection, 
which refer to the rules and regulations affecting political
decision making.}

\subsubsection*{Production, infrastructure, and economic institution}

Suppose the economy has a unique final good, and 
the representative
firm produces output $Y_t$ according to the following
production function:
\begin{equation}\label{infra.prod.equ1}
Y(t) = F[ K(t), L(t), A(t), G(t)],
\mbox{ or } y(t):=f(t,k(t))=F_i[k(t), 1, A(t), G(t)], 
\end{equation}

\noi where $G(t)$ is the aggregate stock of public
goods (or infrastructure) available to all firms at time
$t$. The public good $G(t)$ is a common external 
input to each firm's production function. The government
impose a tax rate $\tau\in [0, 1)$ on output. Hence,
equation for capital accumulation per capita is given by
\begin{equation}\label{infra.per.capital.equ1}
\dot{k}(t) = (1-\tau)f(t, k(t)) - (\delta+ \pi) k(t) - c(t).
\end{equation}

\noi Accordingly, the accumulation
of public goods is described by 
\begin{equation}\label{infra.infra.equ1}
\dot{G}(t)= \tau f(t,k(t)), 
\end{equation}

\noi and firms hire capital and labor to maximize 
$(1-\tau) f(t, k(t)) - w(t)-R(t) k(t)$.

\subsubsection*{Economic structures}

In addition to the components discussed in the previous 
sections, the economic structure of the economy includes 
structures (functionals) of infrastructure investment
and economic institutions. We characterize the level of 
infrastructure investment and its structure by
$(\mathcal{I}, \mathscr{I})$, where $\mathcal{I}=\{ G(t) \} 
\in \mathbbm{R}$ and $\mathscr{I}:=\{
G(\cdot)$ satisfies \eqref{infra.infra.equ1}$\}$.
Infrastructure investment is determined by the
tax rate $\tau$, and its level and structure can be
described by the pair $(\mathcal{T}, \mathscr{T})$,
where the tax level $\mathcal{T}=\{ \tau f(t,k(t)) \}$ 
and the structure (or the functional) of the tax 
$\mathscr{T} :=\{ \tau f(\cdot, \cdot) \}$.
Provided $(\mathcal{I}, \mathscr{I})$ and
$(\mathcal{T}, \mathscr{T})$, the production structure 
$(\mathcal{Y}, \mathscr{Y})$ must be modified 
accordingly. Thus, activities in the economy can be
summarized as
$\mathcal{E}:=( \mathcal{H}, \mathcal{F}, \mathcal{M},
\mathcal{Y}, \mathcal{L}, \mathcal{K}, \mathcal{A},
\mathcal{P}, \mathcal{C}, \mathcal{U}, \mathcal{I}, 
\mathcal{T})$ and its structure can be represented by
$\mathscr{E}:=( \mathscr{H}, \mathscr{F}, \mathscr{M},
\mathscr{Y}, \mathscr{L}, \mathscr{K}, \mathscr{A},
\mathscr{P}, \mathscr{C}, \mathscr{U}, \mathscr{I},
\mathscr{T})$.
Then, given a set of specific structures of 
infrastructure investment and economic institution 
$\{ \mathscr{I}_i, \mathscr{T}_i \}$, $(i\in \mathbbm{I})$
at an initial time, 
the discussion on the social planner's maximization 
problem and its solution is analogous to that in the 
previous sections and hence is skipped here. 

\subsubsection*{Economic implications}

In constrast to infrastructure that can be represented
as public goods produced via explicit production 
functions, most soft infrastructure cannot be described
in this way. However, when the mechanism of the impact of 
soft infrastructure on economic activities can be 
described via explicit functions, the EST approach can 
be extended to characterize the transformation of soft
infrastructue and the corresponding competitive equilibrium
and optimal infrastructure. This would help us
understand the diversity of certain soft infrastructure
in different countries. Take a country's financial
structure as an example. Modern financial institutions,
such as the stock market, venture capital, corporate bonds, 
and large banks have the functions of mobilizing large 
amounts of capital and/or diversifying risks.
The industrial structure of developed countries consists 
of large-scale capital intensive industries that rely on 
risky R\&D for achieving endogenous technological innovation.
Hence, modern financial institutions are appropriate for 
serving the needs of the real sector in developed countries.
However, such financial institutions may not be appropriate 
for developing countries, due to the differences in the
structures of industries and technological progress
between developing and developed countries. 
Since developing countries possess mostly small-scale, 
labor-intensive industries and rely on the adoption of 
mature technologies for technological progress, 
small local or community-based financial institutions 
may be better suited (Lin, 2011).

For other types of soft infrastructure or institutions, 
such as economic policies, protection of intellectual 
property, and other regulations and rules, the EST 
framework can embed them and their transformation into
an economy's development process, which provides
the following implications for studies of economic 
policies and institutions.

First, most studies of economic policies focus on 
the design, implementation, and evaluation of 
specific policies and seldom deal with their dynamic 
changes. As the time scales of economic policies (or 
institutions) are usually larger than those of economic
variables and even some economic substructures, 
transformations of economic  institutions 
and their competitive equilibria can be characterized 
via the EST framework. In particular, as many models in the 
neoclassical sense concentrate on static and 
dynamic equilibria under a given economic policy or 
institution, the EST framework characterizes the 
structural equilibrium of policies and institutions 
in the development process. 

Second, the transformation or birth and decay process 
of a specific economic institution has been largely 
discussed by economic historians, but it has not 
been studied via theoretical models. When different 
development stages of a particular economic  
institution are modeled as different economic structures, 
the EST framework helps us understand the structures of 
economic policies and institutions and their transformations,
for example, the optimal entry and exit times of specific 
policies and institutions.

Third, recent studies show that economic policies can be 
categorized as structural and nonstructural policies
(Abdel-Kader, 2013). Since structural and nonstructural
policies serve different purposes in an economy's 
development process, it is difficult to study both 
types of policies with the existing economic models. The 
EST framework can overcome this difficulty by modeling 
structural and non-structural policies as different
institutional structures and characterizing transformations
among structural and nonstructural policies.

\subsection{Political regimes and institutions}

Another related and interesting issue is the impact of 
political institutions on economic development.
As institutions have different meanings in
different contexts, we define political institutions as 
a system of laws on the organizational form and methods of 
political decision making, such as the organization of 
state power, structural form of the state, political party 
system, and so on. It is not difficult to see that the 
structures of an economy's political institutions still 
have the attributes of durationality and transformality.
Furthermore, political institutions change on a time scale 
much longer than that of economic activities and structures. 

To provide an illustration of the transformation of political
institutions via the EST framework, we extend the discussion
as follows. Suppose that at initial time $t$, the initial 
institutional structure and initial economic structure of 
the country are $\mathscr{N}_{i_1}$ and $\mathscr{E}_{i_2}$, 
respectively, and the initial capital intensity is $k$. 
The dynamic equation for capital intensity under the
given institutional and economic structures $(\mathscr{N}_{i_1}, \mathscr{E}_{i_2})$ may be given by 
$$
\dot{k}(t) = f_{ \mathscr{N}_{i_1}, 
\mathscr{E}_{i_2}}(t,k) - (\delta + \pi) k - c(t).
$$

\noi The total utility and social planner's objective are 
still defined by equations \eqref{sec6.totalutil.equ1} and 
\eqref{sec6.totalutil.equ2}. The competitive equilibrium of 
the economy can be defined similarly as in earlier sections.
Then the solution to the social planner's maximization problem 
satisfies the following.

\begin{proposition}
{\it For each institutional structure $\mathscr{N}_{i_1}
\in \mathcal{N}$ and each economic structure $\mathscr{E}_{i_2}
\in \mathcal{E}$, the value function 
$V_{ \mathscr{N}_{i_1}, \mathscr{E}_{i_2}}(t,k)$ is a
unique viscosity solution of the HJBQVI system }
\begin{equation}\label{sec6.insti.hjbqvi}
\begin{aligned}
& \max \Big\{ \sup_{c \in \mathcal{U}} \Big[ 
\frac{\partial V_{ \mathscr{N}_{i_1}, 
\mathscr{E}_{i_2}}}{\partial t} (t,k) - \big[ f_{ \mathscr{N}_{i_1}, 
\mathscr{E}_{i_2}}(t,k) - (\delta + \pi) k - c(t)\big]
\frac{\partial V_{ \mathscr{N}_{i_1}, 
\mathscr{E}_{i_2}}}{\partial k} (t,k) + e^{-\rho t} u(c) \Big], \\
& \hspace{0.5cm} 
\max \Big[ \max_{\mathscr{E}_{j_2} \neq \mathscr{E}_{i_2} 
}\{ V_{ \mathscr{N}_{i_1}, \mathscr{E}_{j_2}}(t,k)  \}, 
\max_{\mathscr{N}_{j_1} \neq \mathscr{N}_{i_1} } \{ 
V_{ \mathscr{N}_{j_1}, \mathscr{E}_{j_2}}(t,k) \} \Big]
- V_{ \mathscr{N}_{i_1}, \mathscr{E}_{i_2}}(t,k) \Big\}=0.
\end{aligned}
\end{equation}
\end{proposition}

The static and dynamic equilibria of the economy are still 
characterized by the supremum in \eqref{sec6.insti.hjbqvi},
and the structural equilibrium is accounted for by the
second line of the equation. Then, using an argument
analogous to that for Proposition 5.2, we obtain the 
following.
\begin{proposition}
{\it The optimal institutional and economic structures 
at any given time $t$ and with any given capital intensity 
$k$ are characterized by the structural equilibrium 
\eqref{sec6.insti.hjbqvi} at $(t, k)$. Moreover, the 
optimal institutional and economic structures are given 
functions of $t$ and $k(t)$ and, hence, endogenous to the 
capital intensity (or more generally, the factor
endowments of the economy) at time $t$.}
\end{proposition}

We next briefly discuss the implications of equation 
\eqref{sec6.insti.hjbqvi} for the transformation of 
institutional structures in different cases. 
The first case is for developed countries and is 
relatively simple. 
Since developed countries are mostly industrialized, and 
their income per capita is usually higher than that in
developing countries, the social planner (or the economic 
and political elites) in developed countries will not consider 
it necessary to transform their political institutions. 
Or equivalently, the social planner in developed countries 
may have compared different types of institutional structures 
but concludes that the country's current 
institutional and economic structures are better than 
those in other countries. In such case, the structural 
equilibrium of equation \eqref{sec6.insti.hjbqvi}
is not necessary and \eqref{sec6.insti.hjbqvi} 
degenerates to the equation of the supremum, which describes 
the static and dynamic equilibria of the economy with 
fixed institutional and economic structures.

The second case is developing countries focusing on 
``economic development" or ``economic transition" and trying 
to figure out the optimal economic and political institutions 
for the economy. The structural equilibirum in equation 
\eqref{sec6.insti.hjbqvi} is 
\begin{equation}\label{sec6.insti.hjbqvi.equ2}
\begin{aligned}
& \max \Big[ \max_{\mathscr{E}_{j_2} \neq \mathscr{E}_{i_2} } 
\{ V_{ \mathscr{N}_{i_1}, 
\mathscr{E}_{j_2}}(t,k)  \}, 
\max_{\mathscr{N}_{j_1} \neq \mathscr{N}_{i_1}
}\{ V_{ \mathscr{N}_{j_1}, 
\mathscr{E}_{j_2}}(t,k)  \} \Big]
- V_{ \mathscr{N}_{i_1}, \mathscr{E}_{i_2}}(t,k)=0.
\end{aligned}
\end{equation}

\noi This implies that, for the social planner of the economy, 
there are two types of structural transformation
involving institutional and economic structures.
The first type is that the economic structure 
$\mathscr{E}_{i_2}$ transforms into another economic 
structure, while the political institution $\mathscr{N}_{i_1}$ 
remains the same, that is,
\begin{equation}\label{sec6.insti.hjbqvi.equ3}
\max_{\mathscr{E}_{j_2} \neq \mathscr{E}_{i_2} } 
\{ V_{ \mathscr{N}_{i_1}, \mathscr{E}_{j_2}} (t,k)\} = V_{ \mathscr{N}_{i_1}, \mathscr{E}_{i_2}}(t,k).
\end{equation}

\noi The second type is that political institution 
$\mathscr{N}_{i_1}$ and economic structure $\mathscr{E}_{i_2}$
transform into other structures in $\mathcal{N}\times
\mathcal{E}$, which is characterized by
\begin{equation}\label{sec6.insti.hjbqvi.equ4}
\max_{\mathscr{N}_{j_1} \neq \mathscr{N}_{i_1}
}\{ V_{ \mathscr{N}_{j_1}, \mathscr{E}_{j_2}}(t,k)  \}
= V_{ \mathscr{N}_{i_1}, \mathscr{E}_{i_2}}(t,k)
\end{equation}

\noi Equation \eqref{sec6.insti.hjbqvi.equ3} corresponds 
to a scenario in which a developing country develops the 
economy by reforming the economic structures, but the 
political institution remain invariant. By contrast, equation
\eqref{sec6.insti.hjbqvi.equ4} indicates that the 
developing country transforms both the political
institutions and economic structures. 

The choice of equation \eqref{sec6.insti.hjbqvi.equ3} 
or \eqref{sec6.insti.hjbqvi.equ4} for the social planner 
(or the social elite) of the economy to solve 
may lead to different development paths. Take for example
China's and the former Soviet Union's economic transition 
processes. China's economic reform process can be described 
as the process of solving equation \eqref{sec6.insti.hjbqvi.equ3}, whereas the 
economic reform process in the former Soviet
Union and other Eastern European countries can be 
characterized as the process of solving equation 
\eqref{sec6.insti.hjbqvi.equ4}. Although these two processes 
are completely different, they can be described by 
a unified EST framework.

\section{Concluding remarks}

Structural transformation has been discussed intensively
in the literature on economic growth and development over
the past decades. Although sectoral structural transformation
models have been developed to study changes in numerical
economic variables across different sectors, a general 
theoretical framework is still missing 
to characterize a country's full process of 
structural transformation.

The EST framework proposed in this paper bridges this gap 
and makes the following contributions. First, three 
fundamental attributes of structures---struaturality, 
durationality, and transformality---are summarized from 
empirical observations and the literature of economic history. 
Second, with the necessary assumptions on the information 
set of economic structures, a theoretical framework 
is proposed to model the dynamics of economic activities 
and their structures in different time scales and 
characterize the endogenous transformation process of 
structures. Third, we solve the social planner's
optimization problem in the EST model and establish the 
associated competitive equilibrium theory. 
We show that, in addition to the static and dynamic equilibria 
that constitute competitive equilibrium in neoclassical 
growth models, competitive equilibrium in the EST framework
suggests the existence of a third type of equlibrium, 
the structural equilibrium. To demonstrate the 
flexibility of the proposed EST framework, we have discussed
extensions of the EST framework that deal with hierarchical 
production structures, composite structures of consumer 
preference, technological structures via adoption and R\&D, 
changes in infrastructure and economic institutions, 
and switching of political institutions. 

The EST framework provides a method to model the structural 
differences and endogeneity of those differences for 
countries at different levels of development and sheds new 
light on many interesting and oftentimes debated issues. 
We consider a few examples. The import-substitution strategy 
failed in most developing countries in the 1950s and 1960s,
despite the coordination provided by their governments' 
big pushes (Murphy, Shleifer, and Vishny, 1989). The
failure was due to the industrial structure targeted in 
the strategy being too capital intensive while capital was 
scarce in the countries. The growth driven by capital 
accumulation without total factor productivity in 
Singapore and other East Asian economies in their catching-up 
stage was sustainable instead of being doomed to fail, 
as predicted by Krugman (1994). This was because capital 
accumulation is required for upgrading industrial structure 
and technology adoption in the process of catching up and 
returns to capital will not diminish before they reach the 
global production possibility/technology frontier and 
switch to technological innovation through R\&D and growth 
driven by total factor productivity. From the perspective
of EST, the poverty trap or middle-income trap for many 
developing countries is a result of their inability to 
implement dynamic structural transformation due to the 
lack of sufficient capital accumulation to cross the 
capital thresholds required for structural transformation, 
or the lack of government facilitation to overcome coordination 
failures to make the required improvements in hard and soft 
infrastructure for the transformation. 

The EST framework in this paper can be extended to 
incorporate other types of macroeconomic models with 
specific structures, such as models of heterogeneous 
households, heterogeneous firms, overlapping generations, 
trade structures in an open economy, stochastic growth
and so forth. Mathematically, as long as the economic problem 
involves different types of structures, an extended version 
of the EST model can be obtained. Economically, the extended 
EST model and associated competitive equilibrium can be 
characterized by the mathematical method and theory 
developed here, and hence a structural equilibrium 
can be obtained for the extended EST problem. In addition,
since the proposed EST framework can integrate different 
types of economic structures and turn a macroeconomic model 
with a single type of economic structures into one with 
multiple economic structures, the EST framework can integrate 
models of economic development and growth at different 
development stages into a stagewise development and growth 
model. Hence, the EST model provides a unified framework
to account for a country's development and growth process
with structural transformation. 

The mechanism of structural transformation characterized in 
the EST framework assumes that the market is complete, 
information can be freely obtained, and the transformation 
is frictionless, which are certainly not true in the real 
world. Instead, in the real world, different kinds of market 
and information incompleteness and different structures 
require different hard infrastructure and economic as well 
as political institutions to unleash its economic potentials 
in practice. The EST cannot occur spontaneously. Instead, 
policy intervention is usually 
needed to facilitate the transformation of economic structures
(Lin, 1989), which is a topic for further discussion in subsequent research.

\addcontentsline{toc}{section}{References}
\section*{References}
\begin{enumerate}
\item Abdel-Kader, K. (2013) What are structural policies? 
{\it Finance and Development}. 50, 46-47.

\item Acemoglu, D. (2009). {\it Introduction to Modern
Economic Growth}. Princeton, NY: Princeton University Press. 

\item Acemoglu, D. and Guerrieri, V. (2008). 
Capital deepening and non–balanced economic growth. 
{\it Journal of Political Economy}, 116, 467-498.

\item Aghion, P. and Howitt, P. (1992). A model of growth through
creative destruction. {\it Econometrica}, 60, 323-351. 


\item Bardi, M. and Capuzzo-Dolcetta, I. (1997). {\it Optimal 
Control and Viscosity Solutions of Hamilton-Jacobi-Bellman
Equations}. Berlin, Germany: Birkh\"{a}user.





\item Crandall, M. G., Ishii, H., and Lions, P.-L. (1992). A 
user's guide to viscosity solutions, {\it Bulletin of the American
Mathematical Society}, 27, 1-67.

\item Crandall, M. G. and Lions, P.-L. (1984). Viscosity solutions
of Hamilton-Jacobi equations. {\it Transactions of the American
Mathematical Society}, 277, 1-42. 




\item Fleming, W. H. and Soner, H. M. (2006).
{\it Controlled Markov Processes and Viscosity Solutions}.
New York: Springer. 




\item Herrendorf, B., Rogerson, R., and Valentinyi, A. (2014). Growth and
structural transformation. In {\it Handbook of Economic Growth}, Volume 2,
edited by P. Aghion and S. N. Durlauf, 855-941.
Oxford: North Holland.

\item Hirschman, A. O. (1958). {\it The Strategy of Economic 
Development}. New Haven, CT: Yale University Press.

\item Ju, J., Lin, Y., and Wang, Y. (2015).  
Endowment structures, industrial dynamics, and economic
growth. {\it Journal of Monetary Economics}, 76, 244-263.



\item Krugman, P. (1994). The myth of Asia’s miracle. 
{\it Foreign Affairs}, 73 (6), 62-78.

\item Kuznets, S. (1966). {\it Modern Economic Growth}.
New Haven, CT: Yale University Press.

\item Kuznets, S. (1973). Modern economic growth: Findings 
and reflections. {\it American Economic Review}, 63, 247-258.


\item Lewis, W. A. (1954). {\it Economic development with 
unlimited supplies of labor}, Manchester School, 22, 139-191.

\item Lin, J. Y. (1989). An economic theory of institutional change: Induced 
and imposed change. {\it Cato Journal}, 9 (1), 1-33. 

\item Lin, J. Y. (2011). New structural economics: A 
framework for rethinking development. {\it World Bank 
Research Observer}, 26 (2), 193-221.




\item Marx, K. and  Engels, F. (1848). Manifesto of the
Communist Party. {\it Marx/Engels Selected Works}, Vol. 1, Progress Publishers, Moscow, 1969, pp. 98-137.

\item Murphy, K. M., Shleifer, A., and Vishny, R. W. (1989). Industrialization and the big push. {\it Journal of 
Political Economy}, 97 (5), 1003-1026.

\item North, D. C. (1981). {\it Structure and Change in 
Economic History}. New York: W. W. Norton and Company.

\item Ranis, G. and Fei, J. C. H. (1961). A theory of economic development. 
{\it American Economic Review}, 51, 533-565. 

\item Romer, P. M. (1990). Endogenous technological change. {\it Journal 
of Political Economy}, 98, 71-102. 

\item Weber, M. (1904, 1905). {\it The Protestant 
Ethic and the Spirit of Capitalism}, T. Parsons (trans.), 
A. Giddens (intro). London: Routledge, 1992.

\item Xing, H. (2021). Combined Control of Structural 
Transformation and Resource Allocation in Economic 
Dynamical Systems. Working Paper. Department of 
Applied Mathematics and Statistics, Stony Brook University.

\item Xing, H. (2023a). Optimal growth and endogenous switching
of production functions in catching-up economies. 
Working Paper. Department of 
Applied Mathematics and Statistics, Stony Brook University.

\item Xing, H. (2023b). Endogenous growth with human 
capital in catching-up economics. Working Paper. Department of 
Applied Mathematics and Statistics, Stony Brook University.

\item Xing, H. (2023c). Optimal growth and endogenous 
switching of production functions in catching-up economies
with government spending. Working Paper. Department of 
Applied Mathematics and Statistics, Stony Brook University.

\item Xing, H. (2023d). Structural transformation and structural
switching in multi-sector catching-up economies. Working Paper. 
Department of Applied Mathematics and Statistics, 
Stony Brook University.


\end{enumerate}

\end{document}